\documentclass[11pt]{article}

\usepackage{euscript}
\usepackage{amssymb}
\usepackage{amsfonts}
\usepackage{amsbsy}
\usepackage{amsmath}
\usepackage{epsfig}
\usepackage{amsthm}
\usepackage{amscd}
\usepackage{amstext}
\usepackage{verbatim}

\usepackage[pdftex]{hyperref}

\def\ben{\begin{equation}}
\def\een{\end{equation}}
\def\half{{\textstyle{1\over2}}}

\let\a=\alpha \let\b=\beta \let\g=\gamma \let\d=\delta \let\e=\varepsilon
   \let\k=\kappa
\let\l=\lambda

\let\w=\omega \let\G=\Gamma \let\D=\Delta

\let\pa=\partial
\def\be{\begin{equation}}
\def\ee{\end{equation}}
\def\beq{\begin{equation}}
\def\eeq{\end{equation}}
\def\ba{\begin{array}}
\def\ea{\end{array}}

\def\dalemb#1#2{{\vbox{\hrule height .#2pt
       \hbox{\vrule width.#2pt height#1pt \kern#1pt
               \vrule width.#2pt}
       \hrule height.#2pt}}}

\newcommand{\bea}{\begin{eqnarray}}
\newcommand{\eea}{\end{eqnarray}}

\def\ep{{\e}}

\def\R{{{\Bbb R}}}

\def\Lag{{\mathcal{L}}}

\begin{document}

\begin{center}

{ \LARGE {\bf Stellar spectroscopy: \\ Fermions and holographic Lifshitz criticality}}

\vspace{1.2cm}

Sean A. Hartnoll$^\sharp$, Diego M. Hofman$^\sharp$ and David Vegh$^\flat$

\vspace{0.9cm}

{\it $^\sharp$ Center for the Fundamental Laws of Nature, Harvard University,\\
Cambridge, MA 02138, USA \\}

\vspace{0.5cm}

{\it $^\flat$ Simons Center for Geometry and Physics, Stony Brook University, \\
Stony Brook, NY 11794-3636, USA \\}

\vspace{0.5cm}

{\tt hartnoll/dhofman@physics.harvard.edu, dvegh@scgp.stonybrook.edu} \\

\vspace{1.6cm}

\end{center}

\begin{abstract}

Electron stars are fluids of charged fermions in Anti-de Sitter spacetime. They are candidate holographic duals for gauge theories at finite charge density and exhibit emergent Lifshitz scaling at low energies. This paper computes in detail the field theory Green's function $G^R(
\w,k)$ of the gauge-invariant fermionic operators making up the star. The Green's function contains a large number of closely spaced Fermi surfaces, the volumes of which add up to the total charge density in accordance with the Luttinger count. Excitations of the Fermi surfaces are long lived for $\w \lesssim k^z$. Beyond $\w \sim k^z$ the fermionic quasiparticles dissipate strongly into the critical Lifshitz sector. Fermions near this critical dispersion relation give interesting contributions to the optical conductivity.

\end{abstract}

\pagebreak
\setcounter{page}{1}

\tableofcontents

\section{Context}

Known large $N$ gauge theories with holographic gravity duals \cite{Maldacena:1997re} typically have a `microscopic' field content of gauge bosons, scalars and fermions.
These theories often have global $U(1)$ symmetries under which a subset of the scalars and fermions can be charged.
The theory can then be placed at nonzero chemical potential $\mu$ (equivalently, charge density $\langle J^t \rangle$) for this $U(1)$ symmetry.
The dual gravitational description of the nonzero charge density is that there must be an electric flux reaching out to the boundary of the bulk spacetime.
The classical dynamics of the gravity dual will then determine the IR nature of the spacetime subject to this UV boundary condition. This general setup is illustrated in figure \ref{fig:UVIR} below.
\begin{figure}[h]
\begin{center}
\includegraphics[width=230pt]{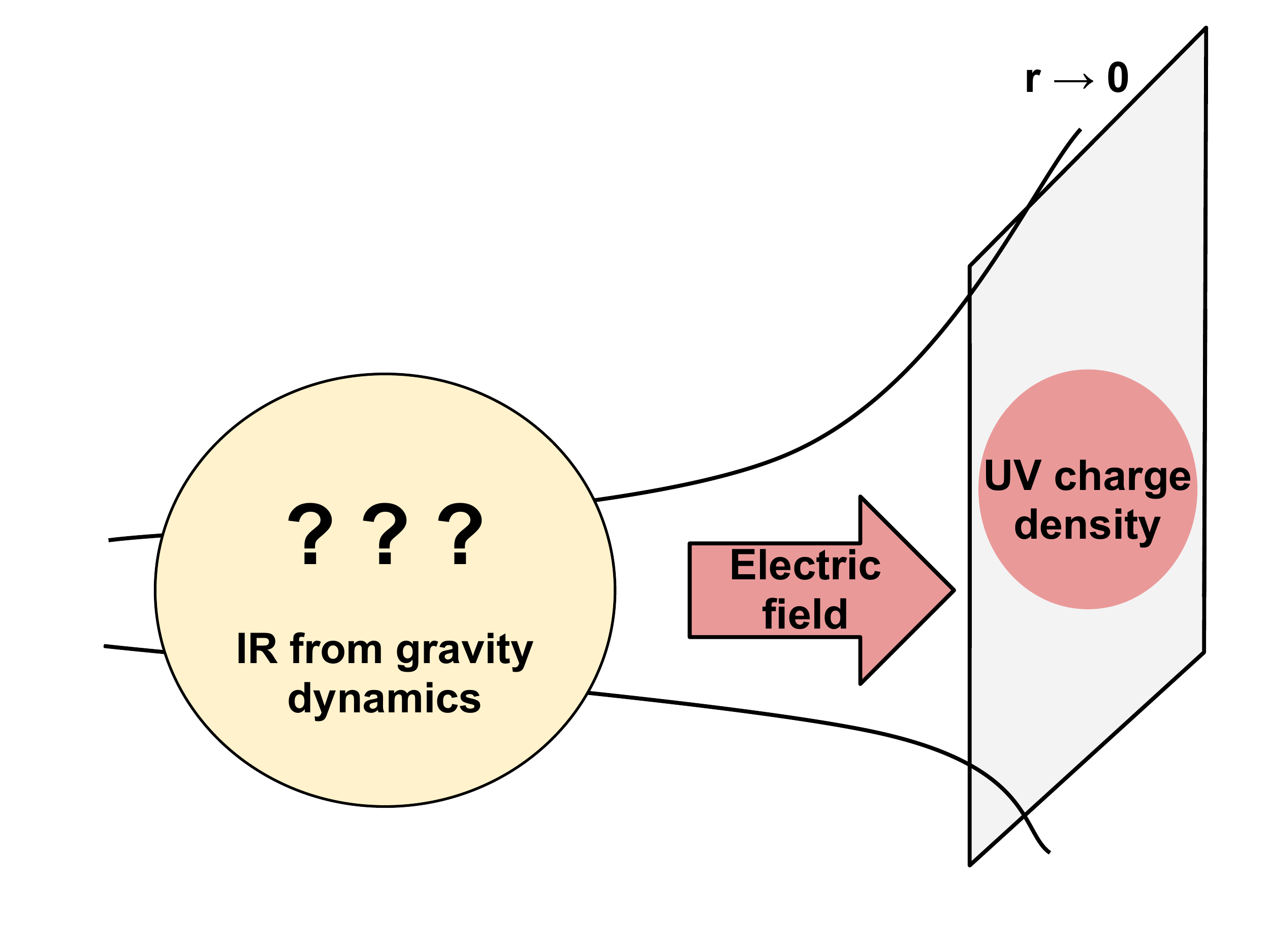}\caption{A nonzero charge density is imposed as a UV boundary condition. The IR physics is to be determined by solving the bulk equations of motion. \label{fig:UVIR}}
\end{center}
\end{figure}

The framework of the previous paragraph has similarities with difficult open questions in the theory of metallic quantum criticality in 2+1 dimensions. There, one typically considers a finite density of itinerant fermions interacting with gapless bosonic degrees of freedom \cite{hertz, review}. The strong effects of forward scattering of the infinitely many excitations at the (putative) Fermi surface invalidate perturbative frameworks for studying the low energy physics of these theories \cite{sungsik, Metlitski:2010pd, Metlitski:2010vm}, although see also \cite{nayak1, nayak2, mross}. Taking the $U(1)$ symmetry of the large $N$ gauge theories as a proxy for the density of fermions (this identification will be literal if there are no massless microscopic bosons charged under the symmetry) we can use the holographic correspondence as a probe of the IR physics of metallically quantum critical states of matter.

The presence of a density of fermions is a defining characteristic of metallic criticality. Fermionic spectral densities are therefore natural observables from which to understand the low energy physics. In particular one is interested in the possible smearing of fermionic quasiparticle poles at a Fermi surface due to interactions with the critical bosonic degrees of freedom. One should bear in mind that the fermions that can be accessed holographically are not the 
microscopic fermions but rather gauge-invariant composite fermionic operators.

The simplest answer to the question posed by figure \ref{fig:UVIR} above, that of determining the IR geometry sourcing an electric flux towards infinity, is the extremal Reissner-Nordstr\"om-AdS black hole \cite{Chamblin:1999tk}. Here there are no charge-carrying fields in the spacetime and the flux emanates from a black hole horizon. The analysis of fermion spectral functions in extremal black hole backgrounds, initiated by \cite{Lee:2008xf}, led to interesting results \cite{Liu:2009dm, Cubrovic:2009ye, Faulkner:2009wj}. The fermion spectral functions exhibit Fermi surface singularities. Furthermore, if the gauge-invariant fermion operator has a sufficiently low scaling dimension relative to its charge, then the pole is broadened into a `non-Fermi liquid' branch cut singularity. The origin of this phenomenon was traced to the behavior of the Dirac equation in the far IR $AdS_2$ geometry of the extremal black hole \cite{Faulkner:2009wj} and can therefore be thought of as indeed being due to strong interactions between fermionic excitations and the quantum critical bosonic modes in the IR.

The chemical potential breaks Lorentz invariance and therefore an emergent scaling symmetry in the IR will generically be characterised by a dynamical critical exponent $z \neq 1$ such that under scaling: $t \to \lambda t$ and $\vec x \to \lambda^{1/z} \vec x$. It is natural to think of such emergent scaling in nonzero density holography as a strong coupling version of the Landau damping of bosons weakly coupled to a Fermi surface \cite{hertz}.
The Reissner-Nordstr\"om-AdS black hole is then seen to be an extreme case of Landau damping in which $z \to \infty$. It is of interest to consider holographic duals that admit, presumably more generic, finite $z$ IR scaling behavior and to characterize the fermionic spectral functions in such theories.
Some general observations about holographic fermions in finite $z$ IR geometries were made in \cite{Faulkner:2010tq}. In particular, it was noted that in a certain regime a Fermi surface pole would not only not be broadened, but that the width should be exponentially small at low energies, making the fermion excitations even more stable than those of a Fermi liquid! We shall reproduce this result, embedded within a slightly broader structure.

Various interesting realisations of finite $z$ IR geometries due to a nonzero charge density in the UV have recently been found. The first were in superconducting phases \cite{Gubser:2009cg, Horowitz:2009ij}. Here the electric field is fully sourced by a condensed scalar field in the bulk, allowing the $AdS_2$ black hole throat to collapse into a `Lifshitz' IR geometry \cite{Kachru:2008yh, Koroteev:2007yp}. We are however interested in non-symmetry broken phases in this paper. A fluid of charged fermions outside of a black hole also absolves the horizon of needing to source the electric field and thereby allows a near horizon Lifshitz geometry \cite{Hartnoll:2009ns}. The resulting geometries were dubbed `electron stars' in \cite{Hartnoll:2010gu} and have many features in common with the superconducting phases just mentioned. However, the $U(1)$ symmetry is not broken by the fermions as these build up a Fermi surface in the bulk rather than forming a coherent state with a definite phase. Finally, without introducing charged matter explicitly, the extremal near horizon $AdS_2$ throat can be collapsed by allowing a dilaton coupling to the Maxwell field in such a way that the effective Maxwell coupling goes to zero at the horizon \cite{Taylor:2008tg, Goldstein:2009cv}. Our computations will focus on the electron star geometries for concreteness, and also because there is a certain minimalism in considering the spectral function of the same fermion that backreacts on the geometry. Nonetheless it will be clear that our results only depend on qualitative features of the geometry and will apply also to the other realisations of finite $z$ we have just mentioned.

Sections 2 and 3 recall the essential features of electron star backgrounds and reduce the computation of the boundary field theory fermion Green's function to a quantum mechanics scattering problem. The fluid description of electron stars is the WKB limit of the bulk Dirac equation, and therefore the rest of the paper unfolds as a WKB analysis of scattering in a specific potential. Many of our results follow qualitatively from an inspection of the effective Schr\"odinger potential, as we discuss in section 3.

Sections 4 through 7 characterise different low energy ($\w \ll \mu$) regimes of the fermion Green's function. The Green's function displays a large number of closely spaced Fermi surfaces with very long lived (as anticipated by \cite{Faulkner:2010tq}) quasiparticle excitations. There is a maximal Fermi momentum $k_F^\star$. While this extremal momentum dominates the quantum oscillations of the star \cite{Hartnoll:2010xj}, we show here it has the smallest spectral weight of all the Fermi surfaces. We show that the sum of volumes of the 2+1 dimensional field theory Fermi surfaces adds up to the total charge density, implying that the Luttinger count holds for this (non-Fermi liquid) system. Excitations of the Fermi surfaces become unstable at small momenta $k \lesssim \w^{1/z}$, where they dissipate efficiently into the critical Lifshitz sector. We characterise carefully the poles in the Green's function close to the critical dispersion $\w \sim k^z$ and estimate that they give a dominant contribution to the optical conductivity of the system.

Finally in section 8 we take a limit in which the entire star is contained within the Lifshitz region of the spacetime. In this limit and with $z=2$ we can solve for the Green's function exactly, thereby giving explicit examples of various statements made throughout the paper.

\section{The electron star background}

This paper will consider the fermionic spectral functions of field theories with electron star gravity duals. As mentioned in the introduction, our results mainly depend on qualitative features of the geometry that are shared by other spacetimes. We will emphasize these aspects throughout.

The background spacetime and Maxwell field take the form
\be\label{eq:metric}
ds^2 = L^2 \left(- f dt^2 + g dr^2 + \frac{1}{r^2} \left( dx^2 + dy^2 \right) \right) \,, \qquad A = \frac{e L}{\k} h dt \,,
\ee
with $\{f,g,h\}$ functions of the radial coordinate $r$. In these coordinates $r \to 0$ at the spacetime boundary.
Throughout, our spacetime will be 3+1 dimensional and the dual gauge theory, 2+1 dimensional.
The equations of motion for the metric functions follow from the Einstein-Maxwell-charged fluid Lagrangian density
\be\label{eq:lagrangian}
\Lag = \frac{1}{2 \k^2} \left(R + \frac{6}{L^2} \right) - \frac{1}{4 e^2} F_{ab} F^{ab} + p(\mu,s) \,.
\ee
as described in \cite{Hartnoll:2010gu}. Here $p$ is the local pressure of a bulk fermion fluid in terms of a local chemical potential and a local entropy density. The fluid is a corse-grained description of free Dirac fermions with unit charge (without loss of generality) and mass $m$. The fluid description is valid when \cite{Hartnoll:2010gu} (analogous in depth discussions are also found in \cite{deBoer:2009wk, Arsiwalla:2010bt})
\be\label{eq:sim}
e^2 \sim \frac{\k}{L} \ll 1 \,,
\ee
and the following dimensionless mass is order one
\be\label{eq:mhat}
\hat m^2 = \frac{\k^2}{e^2} m^2 \,.
\ee
This second condition implies that $L^2 m^2 \sim e^2 L^2/\k^2 \sim 1/e^2 \gg 1$. The Compton wavelength of the fermion is much smaller than the radius of curvature of the spacetime and therefore we should expect to be in a WKB limit. This WKB limit will greatly simplify our discussion of the fermion Green's function below. While convenient, some of the essential physics should not depend strongly on this limit. We hope to explore a `quantum electron star' explicitly in the future.

In the deep IR, $r \to \infty$, the functions $\{f,g,h\}$ are found to tend towards a Lifshitz geometry
\be\label{eq:IRlif}
f = \frac{1}{r^{2z}} \,, \qquad g = \frac{g_\infty}{r^2} \,, \qquad h = \frac{h_\infty}{r^z} \,,
\ee
with in particular \cite{Hartnoll:2010gu}
\be\label{eq:i1}
h^2_\infty = \frac{z-1}{z} \,.
\ee
The value of the dynamical critical exponent $z$ depends on the order one dimensionless parameters $\hat m$ and $e^4 L^2/\k^2$. It is bounded from below by \cite{Hartnoll:2010gu}
\be\label{eq:i2}
z \geq \frac{1}{1 - \hat m^2} > 1 \,.
\ee
This is the finite $z$ emergent IR criticality advertised in the introduction.

Integrating outwards from the IR, the local pressure of the fermion fluid decreases monotonically until it vanishes at a radius $r = r_s$. This is the boundary of the electron star. Outside the electron star, $r < r_s$, the spacetime is that of Reissner-Nordstr\"om-AdS
\be\label{eq:RNA}
f = c^2\left(\frac{1}{r^2} - \hat M r + \frac{r^2 \hat Q^2}{2} \right) \,, \qquad g = \frac{c^2}{r^4 f} \,, \qquad h = c \left(\hat \mu - r \hat Q \right) \,.
\ee
The four constants here $\{c, \hat\mu, \hat M, \hat Q\}$ are determined by matching at $r=r_s$ and in turn determine the thermodynamics of the dual field theory.

\section{The WKB Dirac equation}

Our objective is to compute the boundary Green's function of the gauge-invariant fermion operators that are making up the electron star. The bulk fermion can be taken to have four components and to satisfy the Dirac equation. With a judicious choice of basis for the Gamma functions, the Dirac equation separates into two decoupled equations for two two-component spinors. These equations differ only by the momentum $k \to -k$. Without loss of generality we can focus on one of the two-component spinors. The Dirac equation can then be written \cite{Faulkner:2009wj}
\be\label{eq:dir}
\sqrt{g^{rr}} \sigma_3 \pa_r \Phi - m \Phi - i \sqrt{g^{xx}} k \sigma_2 \Phi + \sqrt{|g^{tt}|} (\w + A_t) \sigma_1 \Phi = 0 \,.
\ee
Here $\Phi$ is a two component spinor, the $\sigma_i$ are Pauli matrices, and we have taken the field theory spacetime dependence $\Phi \propto e^{- i \w t + i k x}$. The field $\Phi$ is rescaled relative to the actual Dirac field: $\Phi = (-g g^{rr})^{\frac{1}{4}} \psi$.

The limits (\ref{eq:sim}) and (\ref{eq:mhat}) force us to consider the WKB limit of the Dirac equation. The large parameter that appears will be
\be\label{eq:gammadef}
\fbox{$\displaystyle \frac{e L}{\k} \equiv \gamma \gg 1.$}
\ee
In particular, we can write the spinor as
\be\label{eq:WKBansatz}
\Phi =
{1 \choose \phi} A \, e^{ i \g S} \,.
\ee
Furthermore, in order to obtain an interesting $\w$ and $k$ dependence, we should rescale
\be\label{eq:wkhat}
\w = \g \hat \w \,, \qquad k = \g \hat k \,.
\ee
Then to leading order in $\g \gg 1$ the Dirac equations become
\bea
\frac{i S'}{\sqrt{g}} + \frac{(h + \hat \w) \phi}{\sqrt{f}} - r \hat k \phi - \hat m & = & 0 \,, \\
- \frac{i S'}{\sqrt{g}} + \frac{(h + \hat \w)}{\sqrt{f} \, \phi} +\frac{r \hat k}{\phi} - \hat m & = & 0 \,.
\eea
These equations are simple to solve. The solutions are
\bea
S & = & \pm \int^r dr \sqrt{g} \sqrt{(\hat \w + h)^2/f - r^2 \hat k^2 - \hat m^2} \,, \label{eq:SS} \\
\phi & = & \frac{\hat m \mp i \sqrt{(\hat \w + h)^2/f - r^2 \hat k^2 - \hat m^2}}{(\hat \w + h)/\sqrt{f} - r \hat k} \,. \label{eq:phiphi}
\eea
The overall function $A$ is easily determined by working to next order in the WKB expansion. The zero temperature `horizon' is at $r \to \infty$. We must impose ingoing boundary conditions at the horizon. The ingoing solution is, for $\w > 0$, the upper of the above signs. However, we must be careful to discuss the matching of the solution across the turning points of the exponent $S$.

In order to match across the turning points it is convenient to map the Dirac equation into a Schr\"odinger form. This will allow us to use the standard machinery from WKB approximations in quantum mechanics. We will thus rederive the expressions (\ref{eq:SS}) and (\ref{eq:phiphi}) from this perspective. Writing
\be
\Phi =
{\Phi_1 \choose \Phi_2} \,,
\ee
it is simple to obtain a second order equation for each component of the spinor. The expression is a little clumsy, but it simplifies in our limit $\g \gg 1$ to
\bea\label{eq:schro}
\Phi_1'' = \g^2 g \left(\hat m^2 + r^2 \hat k^2 - \frac{(\hat \w + h)^2}{f} \right) \Phi_1 \,, \\
 \Phi_2 = \frac{1}{(\hat \w + h)/\sqrt{f} - r \hat k} \left(\hat m \Phi_1 - \frac{1}{\g} \frac{1}{\sqrt{g}} \Phi_1' \right) \,. \label{eq:cons}
\eea
Thus we recover the leading order exponents (\ref{eq:SS}) in the usual WKB fashion. We will proceed shortly to perform a WKB study of the `zero energy' Schr\"odinger equation (\ref{eq:schro}) in the following section, but some comments first.
The first of the two equations above has corrections of order $\g$.\footnote{\label{eq:foot}For completeness, the first order correction to the right hand side of equation (\ref{eq:schro}), the Schr\"odinger potential, is $\g \hat m \sqrt{g} \frac{(\hat \w + h) f'/(2f)+\hat k \sqrt{f} - h'}{(\hat \w + h) - \hat k r \sqrt{f}}$. In order to obtain this expression it is necessary to rescale $\Phi_1$ to eliminate a first derivative term that appears in (\ref{eq:schro}) at this order and thereby maintain the Schr\"odinger form.} These corrections give a contribution to the WKB wavefunction of the same order as the standard $V^{-\frac{1}{4}}$ WKB prefactor, with $V$ the Schr\"odinger potential. That is to say, a correction of order $\g^0$ in the exponent. Such corrections
can have an effect on the phases that appear in performing matchings at points where the potential vanishes. These phases are responsible for the `$\half \pi$' type shifts in the ground state energy. In this paper we will not be overly concerned with the precise energies of the low lying states, but rather in states with intermediate to large excitation number. For these states, the effects of corrections to (\ref{eq:schro}) are negligible (although computable if necessary) and we shall not consider them further. Note however that the final term in the second equation above needs to be kept in the WKB limit as the derivative brings down an extra power of $\g$.

A second comment concerning (\ref{eq:cons}) is that it appears to become singular when the denominator $(\hat \w + h)/\sqrt{f} - r \hat k$ vanishes. The higher order corrections to the potential in (\ref{eq:schro}) in fact also diverge at such points. These divergences are an artifact of eliminating variables and are not inherent to the Dirac equation (\ref{eq:dir}). The easiest way in practice to avoid this complication is to note that if we instead obtain a Schr\"odinger equation for $\Phi_2$ and then obtain $\Phi_1$ from $\Phi_2$, the divergence now occurs at $(\hat \w + h)/\sqrt{f} + r \hat k = 0$ while the effective Schr\"odinger equation remains the same (\ref{eq:schro}). Thus to avoid possible singularities, we can swap between solving for $\Phi_1$ or $\Phi_2$ first if we approach a singular point. We are interested in real $k$ and $r$ only, so nontrivial monodromies should not arise. The upshot is that we can just focus on the Schr\"odinger equation (\ref{eq:schro}). We see that only $k^2$ appears in the effective Schr\"odinger equation and so in the following we will restrict to $k > 0$ without loss of generality.


The nature of the WKB solutions is controlled by the turning points in the exponent. From the turning points alone we will obtain a good picture of the Green's function we are after. Thus, we need to know the sign of the potential appearing in the effective zero energy Schr\"odinger equation (\ref{eq:schro})
\be\label{eq:potential}
\fbox{$\displaystyle V \equiv g  \left( r^2 \hat k^2 +  \hat m^2 - \frac{(\hat \w + h)^2}{f} \right) \,. $}
\ee
In discussing this quantity, it will be useful to introduce the following notions. The local Fermi momentum is
\be\label{eq:KFr}
r^2 \hat k_F^2(r) \equiv \frac{h^2}{f} - \hat m^2 \,.
\ee
The boundary of the star $r=r_s$ is determined by \cite{Hartnoll:2010gu}
\be
\hat k_F(r_s) = 0 \,.
\ee
Note that $ \hat k_F^2(r)$ is negative for $r < r_s$, i.e. outside the star. All the interesting features we are about to explore occur inside the electron star.
The `extremal' local Fermi surface is determined by \cite{Hartnoll:2010xj}
\be\label{eq:KF}
k_F^\star \equiv k_F(r_\star) = \max_{r} k_F(r) \,.
\ee
The extremal radius always exists and is unique with $r_\star > r_s$, inside the star. It can be determined numerically \cite{Hartnoll:2010xj}. The local Fermi momentum $k_F(r)$ goes to zero at the center of the star as well as at the star boundary.

\begin{figure}[h]
\begin{center}
\includegraphics[width=150pt]{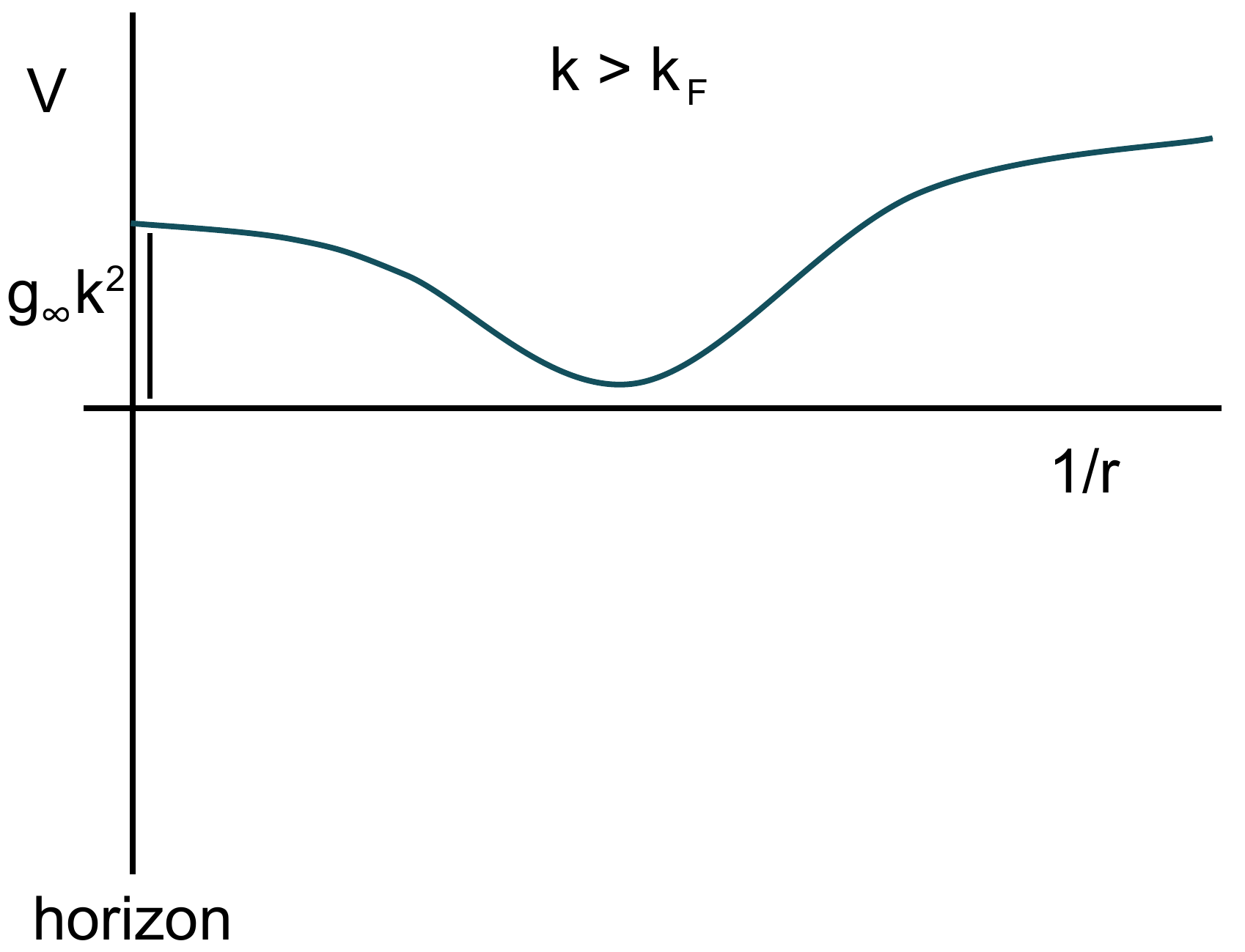}\hspace{2cm}\includegraphics[width=150pt]{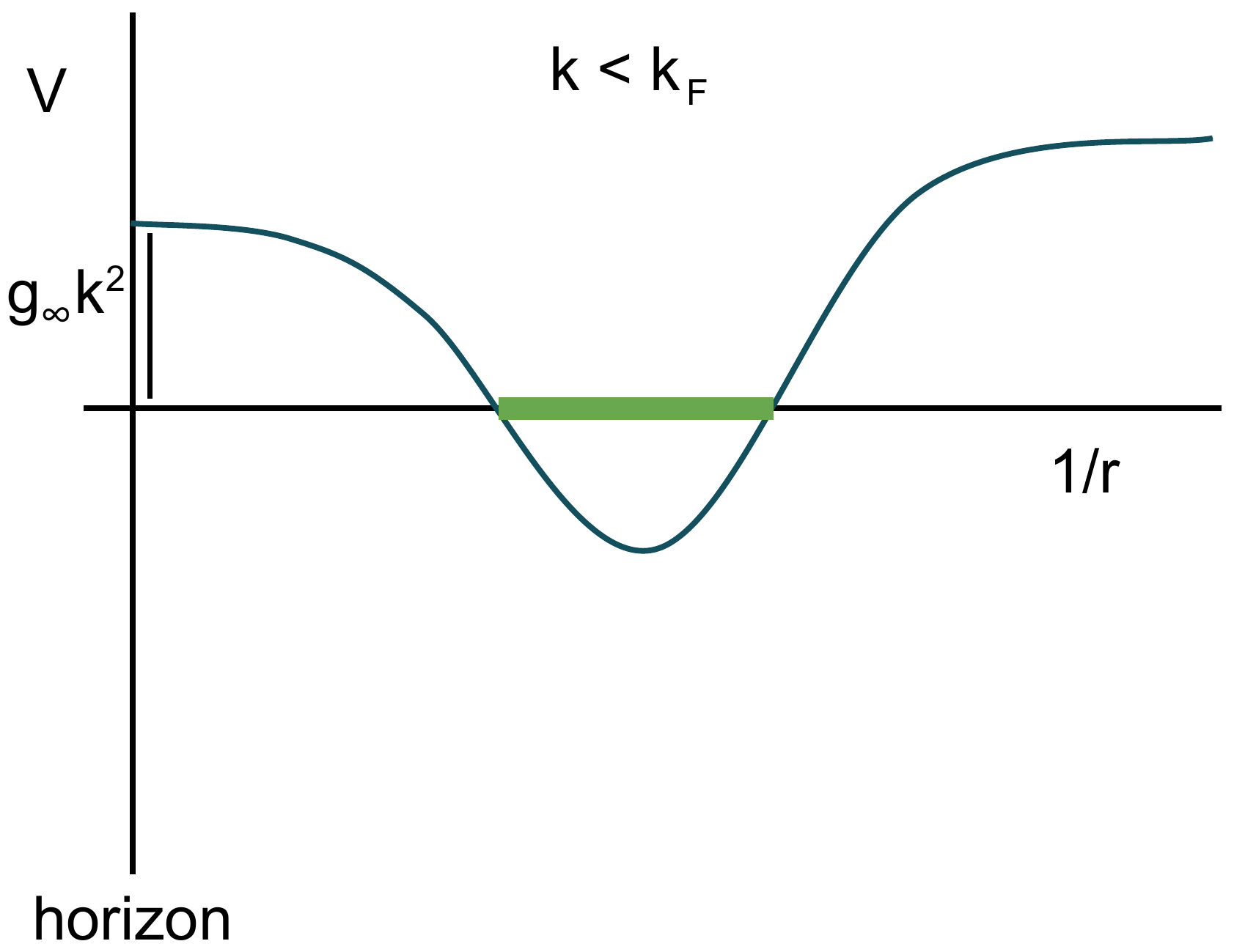}
\caption{Illustrative radial dependence of the Schr\"odinger potential when $\w = 0$. The thick green line denotes oscillatory regions. To the left is the horizon while the boundary is to the right at $r \to 0$. \label{fig:cases0}}
\end{center}
\end{figure}

We are interested in the low frequency Green's function $\w \ll \mu$. Start by setting $\w = 0$. At zero frequency
\be\label{eq:VKF}
V =  g \, r^2 \Big(\hat k^2 -  \hat k_F^2(r)\Big) \,, \qquad (\w = 0) \,.
\ee
Thus we have, for $\w = 0$,
\bea
k > k_F^\star & \Rightarrow & V > 0 \quad \text{everywhere} \,, \label{eq:w0a} \\
k < k_F^\star & \Rightarrow & V < 0 \quad \text{for} \quad r_1 < r < r_2 \quad \text{and} \quad V > 0 \quad \text{elsewhere} \,. \label{eq:w0b}
\eea
Here clearly $r_{1,2}$ are defined as the two solutions to $k = k_F(r)$. 
Thus we see that the zero frequency WKB Dirac solution is exponential wherever the momentum is greater than the local Fermi momentum and oscillating whenever the momentum is lower than the local Fermi momentum. The oscillations therefore occur when the local states with the given momentum are filled. We may expect some resonant behaviour at $k = k_F^\star$ separating the two regimes. The two cases are illustrated in figure \ref{fig:cases0}.

Now consider low but nonzero frequencies $0 < |\w| \ll \mu$. This should only affect the zero frequency results (\ref{eq:w0a}) and (\ref{eq:w0b}) in the near horizon region. This is because the near horizon ($f, h \to 0$) and zero frequency limits do not commute. Away from the infinite redshift of the horizon, the effects of small nonzero $\w$ are uniformly small.

In the near horizon Lifshitz region (\ref{eq:IRlif})
\be\label{eq:nearE}
\fbox{$\displaystyle V_\text{Lif.} = g_\infty \left(\hat k^2 + \frac{\hat m^2 - (h_\infty + r^z \hat \w)^2}{r^2}\right) \,.$}
\ee
In the outer limits of the near horizon region, where $\hat \w r^z \ll 1$, we recover the zero frequency conditions of (\ref{eq:w0a}) and (\ref{eq:w0b}). This is the regime of overlap of the near and far regions and is consistent with our previous remark that small $\w$ should not qualitatively change the turning points in the far region. Note that the relations (\ref{eq:i1}) and (\ref{eq:i2}) imply that in the near horizon region ${\displaystyle \lim_{r \to \infty}} r^2 \hat k_F^2 = h_\infty^2 - \hat m^2 \geq 0$. This is the consistency condition that the interior of the star is in fact occupied by fermions.

At any nonzero frequency we see that, assuming $z>1$, for any mass and momentum, $V < 0$ in the far interior $r \to \infty$. This fact may require another change in sign of $V$ at a radius $r_3$ in the near horizon region and leaves us with three possible patterns of signs shown in the following table. The possibilities are also illustrated in figure \ref{fig:cases} below.
\begin{table}[h]
\begin{center}
  \begin{tabular}{@{} |r|c|c|c|c| @{}}
    \hline
     Case & $r \to \infty$ & $r_3 > r > r_2$ & $r_2 > r > r_1$ & $r \to 0$ \\ 
    \hline
    $I: \quad k > k_F^\star$ & $V < 0$ & \multicolumn{3}{c}{$V > 0$} \vline \\ \hline
    $II: \quad k < k_F^\star$ & $V < 0$ & $V > 0$ & $V < 0$ & $V > 0$ \\ \hline
    $III: \quad k < k_F^\star$ & \multicolumn{3}{c}{$V < 0$} \vline & $V > 0$ \\ 
    \hline
  \end{tabular}
  \caption{Three possible patterns of signs at a small but nonzero frequency $\w$.}\label{tab:three}
\end{center}
\end{table}

For the the third case shown in table \ref{tab:three} to occur, it must be that the second intersection point $r_2$ at $\w=0$ occurred within the near horizon region. This is because the effects of finite $\w$ are negligible outside this region.
\begin{figure}[h]
\begin{center}
\includegraphics[width=150pt]{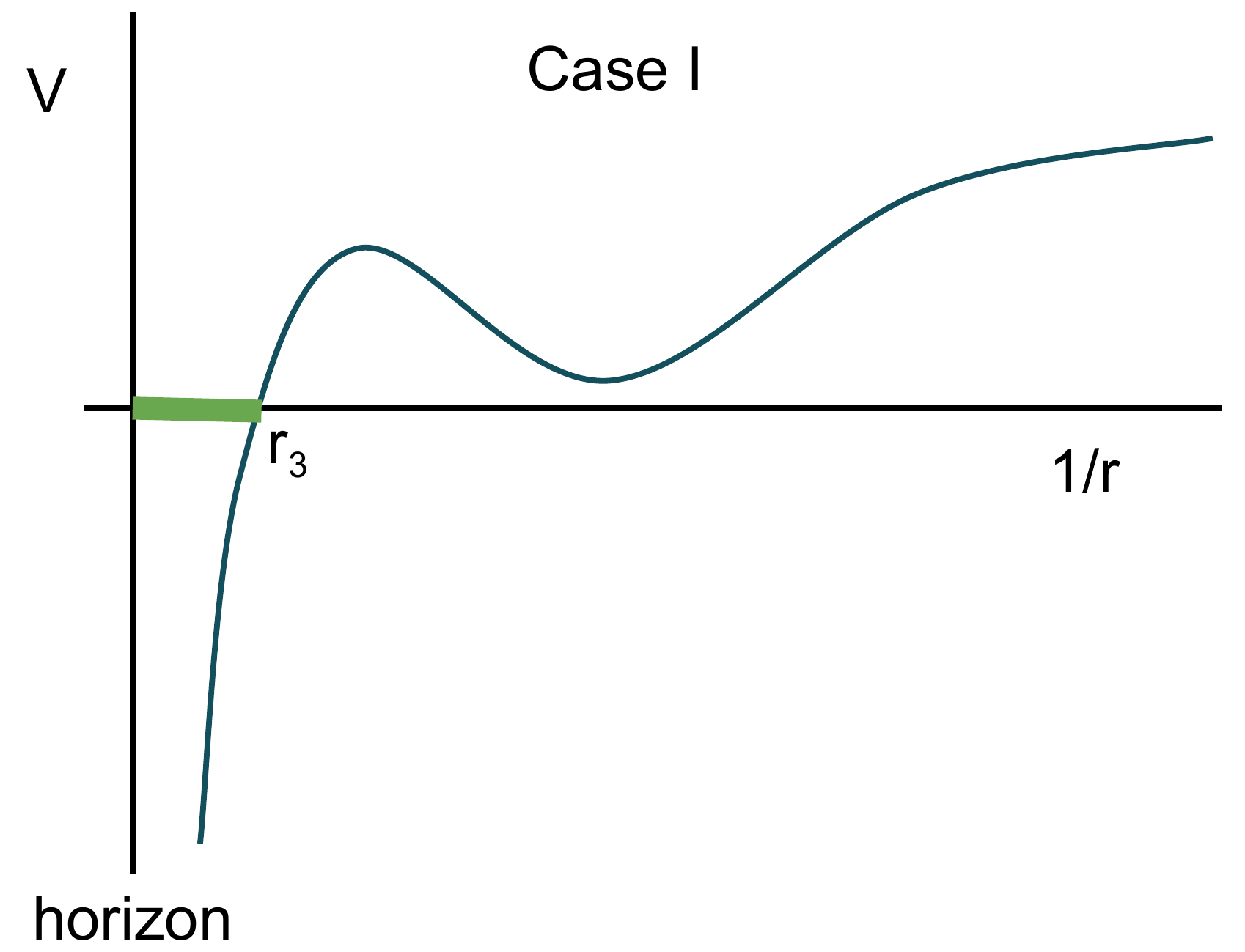}\includegraphics[width=150pt]{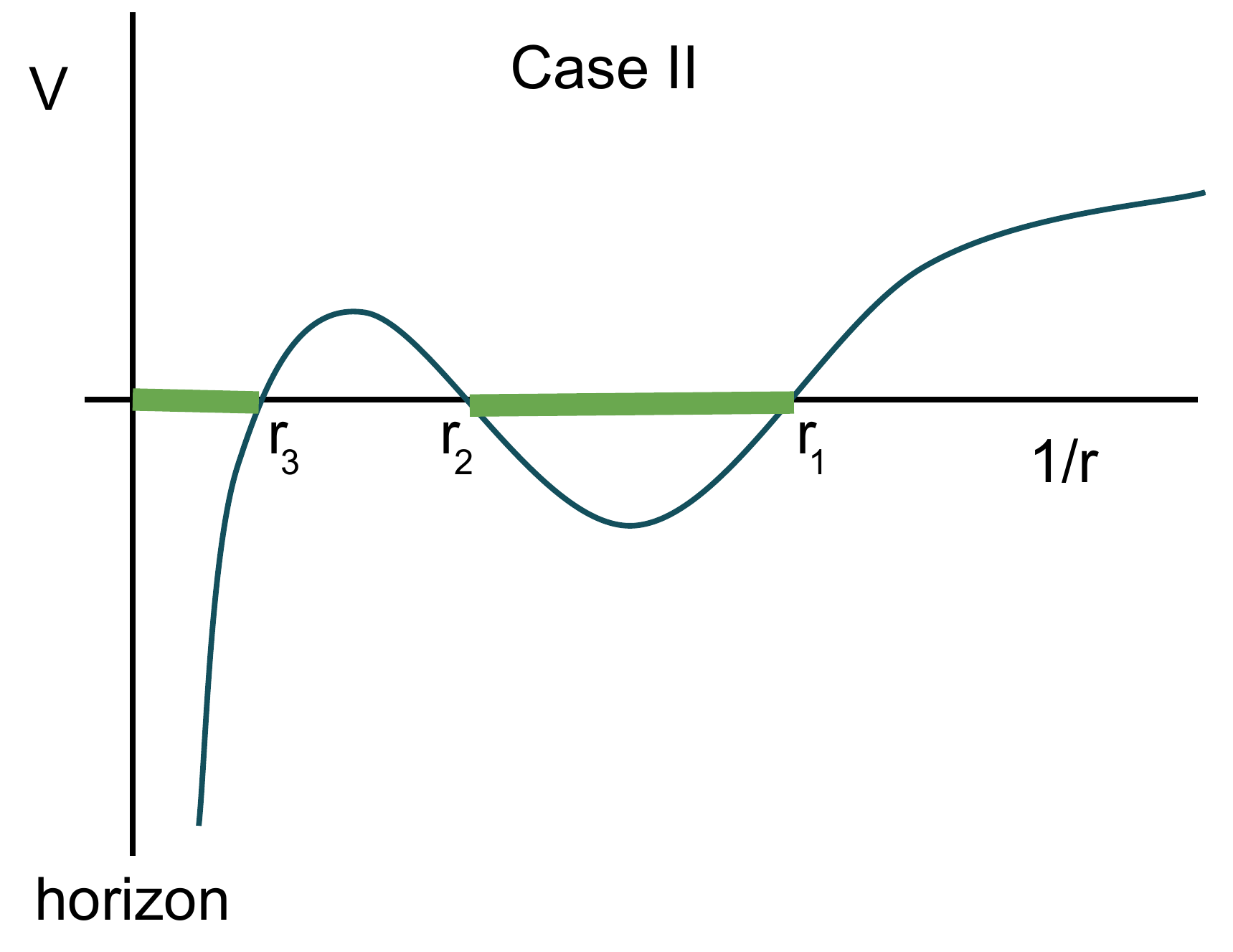}\includegraphics[width=150pt]{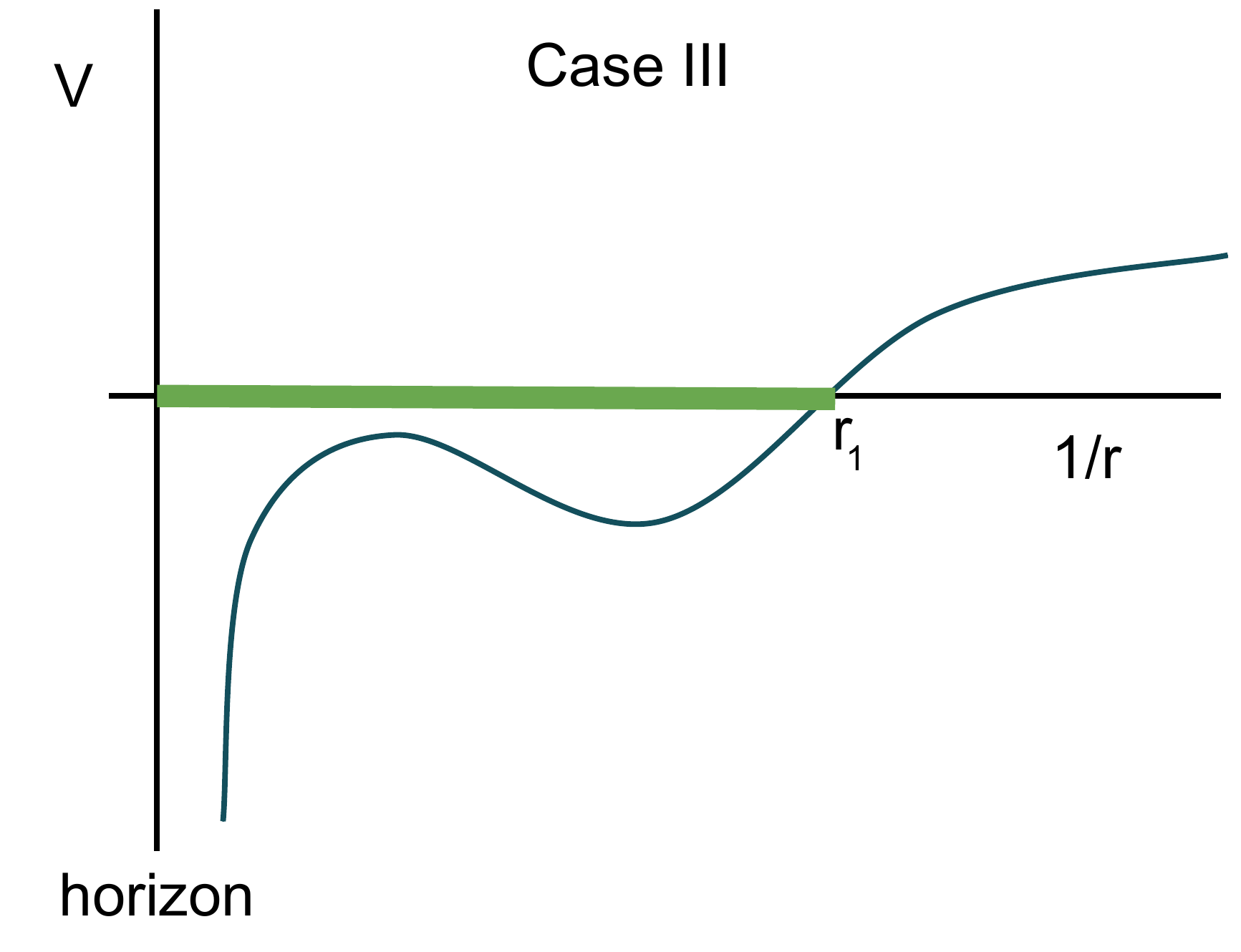}\caption{Illustrative radial dependence of the Schr\"odinger potential (with $z>1$) in the three possible cases at small nonzero frequency. The thick green line denotes oscillatory regions. To the left is the horizon while the boundary is to the right at $r \to 0$. \label{fig:cases}}
\end{center}
\end{figure}
Furthermore, it must be that the momentum $k$ is small, so that its effects in making $V$ positive can be offset with a small frequency $\w$. In fact, checking for the existence of zeros of (\ref{eq:nearE}) we can determine precisely when the third case can occur
\be\label{eq:three}
\text{Case III} \qquad \Leftrightarrow \qquad \hat \w > |\hat k|^z \; \max_{0 < x < \infty} \; \frac{\sqrt{\hat m^2 + x^2}-h_\infty}{x^z} \,.
\ee
It is straightforward to find the maximum analytically, and we shall do so in a later section.

In the following sections we will obtain the boundary fermion Green's function in some detail. However, a qualitative understanding of the frequency and momentum dependence of the spectral density -- the imaginary part of the Green's function -- can be extracted simply from the Schr\"odinger potentials of figure \ref{fig:cases}. This gives us a global overview of what to expect in the following.

\begin{figure}[h]
\begin{center}
\includegraphics[width=280pt]{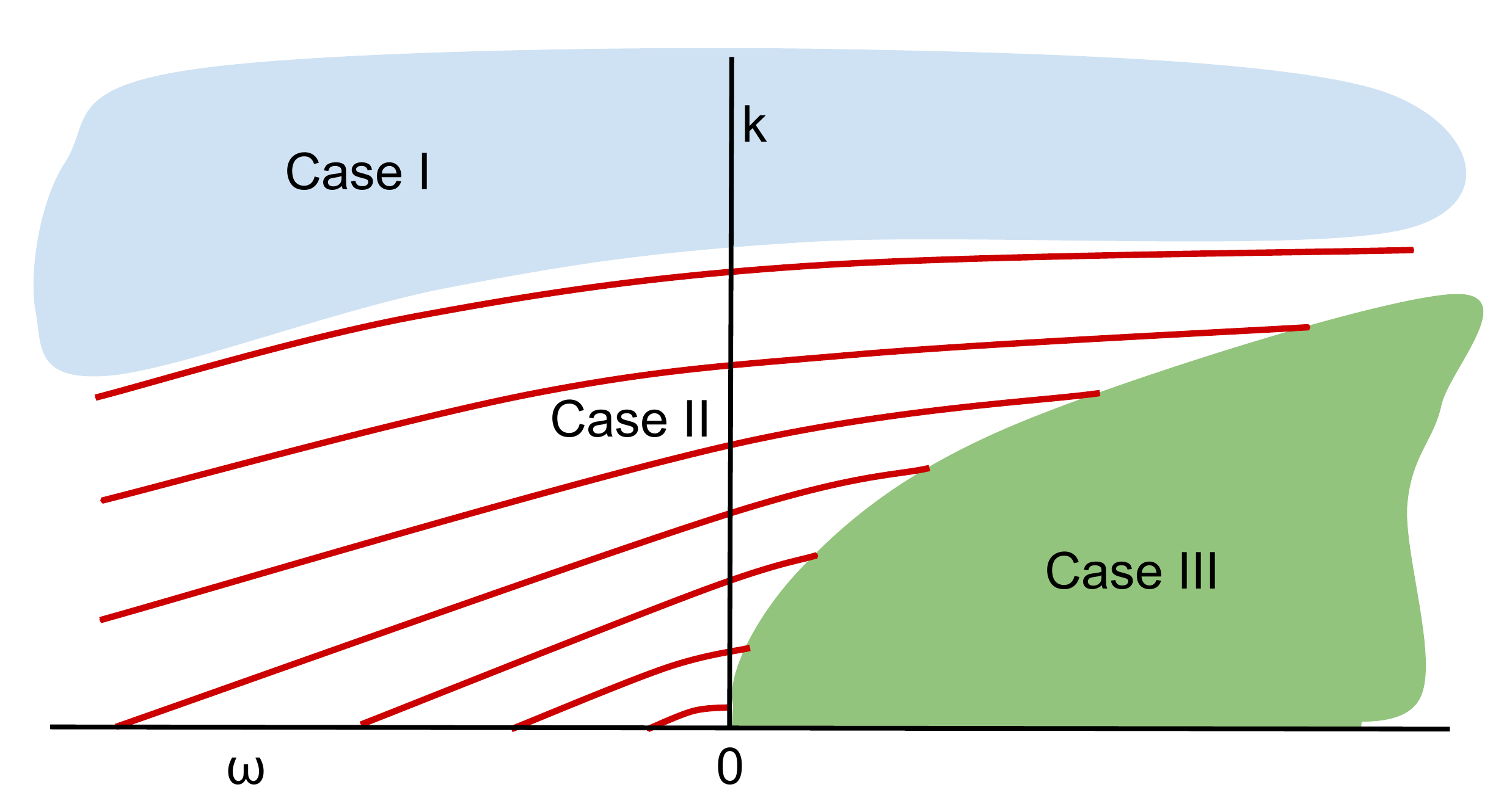}\caption{Schematic dependence of the dual field theory fermion spectral density as a function of $\w$ and $k$ for $z>1$. Red lines denote the dispersion of sharp Fermi surface poles. In case I there are no low energy excitations because all the available Fermi momenta are too far away. In case III there are no well defined excitations because the Fermi surface has become unstable towards dissipation into the critical Lifshitz sector. \label{fig:wkplane}}
\end{center}
\end{figure}

We can see that in case II there will be many (because we are in the WKB limit) almost bound states that are stable up to an exponentially small tunneling amplitude into the horizon. These correspond to many Fermi surface poles in the retarded Green's function. At small frequencies the poles are bounded above in momentum by $k = k_F^\star$ and below by the Lifshitz dispersion $k \sim \w^{1/z}$. These are the boundaries with cases I and III respectively. The boundary regions themselves are particularly interesting and we will consider them carefully below. In cases I and III we should expect no sharp features in the spectral density as there are no almost stable states. Putting these statements together leads to the schematic figure \ref{fig:wkplane}. This plot is very similar to plots describing the fermion spectral density in a superconducting background \cite{Faulkner:2009am, Gubser:2009dt}.
This is unsurprising because the background fluid of charged fermions has many features in common with a background charged bosonic condensate. Furthermore, the backgrounds \cite{Horowitz:2009ij, Gubser:2009gp} considered in \cite{Faulkner:2009am, Gubser:2009dt} also had an emergent ($z=1$) scaling symmetry in the IR and therefore a `light-cone' within which fermion excitations were efficiently dissipated. One special feature of the $z=1$ case is that the lightcone exists for both positive and negative frequencies. In our figure we have a `non-relativistic lightcone' that only exists at positive frequencies. We will discuss the physics underlying figure \ref{fig:wkplane} in the following section. In a later section we will present a non-schematic version of this figure. We will also be concerned with the extent to which the sharp Fermi surfaces can be `smeared' into a Fermi ball \cite{Hartnoll:2010xj} and with the r\^ole of the relative residues of Fermi surface poles. Some of the Fermi surfaces will have a spectral weight that is exponentially suppressed relative to others.

\section{The boundary fermion Green's function: generic $\w, k$}

The Green's function of the fermion operator in the quantum field theory dual to our bulk fermion is read off from the asymptotic behaviour of solutions to the bulk Dirac equation.
Dropping the subscript $1$ for the spinor component, recall that we are interested in solving the Schr\"odinger equation (\ref{eq:schro})
\be\label{eq:S2}
- \Phi'' + \g^2 V \Phi = 0 \,,
\ee
with the potential given by (\ref{eq:potential}). Given that $\g \gg 1$, the WKB approximation will be valid as long as the potential and its first two derivatives remain of order one. This follows from the fact that the WKB approximation requires that the derivatives of the potential are small compared to the potential. We can use dimensional analysis to see how the comparison should be made. Because $V$ has dimension two, we require:
\be\label{eq:wkbconditions}
|V '(r)| \ll \g |V(r)|^\frac{3}{2} \,, \quad \text{and} \quad |V''(r)| \ll \g^2 |V(r)|^2 \,.
\ee
These inequalities are obviously satisfied if $V$ is order one with $\gamma \gg 1$. Whenever $V$ is smaller than order one we should consider a matching strategy over the problematic region. We will come to these cases later.

In the outer region, defined to be the region from the boundary to the first turning point, we have $V > 0$. The two independent solutions of (\ref{eq:S2}) are therefore schematically
\be\label{eq:pm}
\Phi \sim (\g^2 V)^{-\frac{1}{4}} \exp\left\{\pm \g \int_\epsilon^r dr \sqrt{V}\right\} \,.
\ee 
Here $\epsilon \ll 1$ is a UV cutoff. Outside the star we must use the Reissner-Nordstrom-AdS expressions (\ref{eq:RNA}). It is then possible to show that to leading order near the boundary (recall $m L \gg 1$)
\be\label{eq:phi1}
\Phi = a_+(\w,k) r^{m L} + a_-(\w,k) r^{1 - m L}\,.
\ee
To obtain this precise result, including the leading correction to the powers $\pm m L$, one needs to include the leading correction, in footnote \ref{eq:foot} above, to the Schr\"odinger potential as well as the $V^{-\frac{1}{4}}$ prefactor in the WKB expression (\ref{eq:pm}). This is the only place in this paper where we will refer to this footnote and indeed we do not even strictly need this subleading correction here. Imposing ingoing boundary conditions at the horizon and matching across the turning points, we will obtain some combination of these two modes in the outer region. Being careful to solve both (\ref{eq:schro}) and (\ref{eq:cons}) consistently and keeping the leading terms near the boundary $r \to 0$
\be\label{eq:pm1}
\Phi_\text{outer} = A(\w,k) r^{-m L}{0 \choose 1}   + B(\w,k) r^{m L}  {1 \choose 0}  \,.
\ee
The retarded Green's function of the boundary operator dual to $\Phi$ is then (e.g. \cite{Faulkner:2009wj})
\be\label{eq:pm2}
G^R(\w,k) = \frac{B(\w,k)}{A(\w,k)} \,.
\ee
Because we will be focussing on the single Sch\"odinger equation (\ref{eq:S2}) for $\Phi$ it is convenient to re-express the Green's function in terms of the asymptotic behaviour (\ref{eq:phi1}) of this function. Thus, using (\ref{eq:cons}), we obtain
\be\label{eq:aa2}
G^R(\w,k) = \frac{\hat \mu + \hat \w/c - \hat k}{2 \hat m} \, \frac{a_+(\w,k)}{a_-(\w,k)} \,.
\ee
Here $\hat \mu$ and $c$ are the boundary chemical potential and speed of light as appearing in (\ref{eq:RNA}). $\hat \mu$ has slightly unconventional units due to compatibility with the notation in previous works.

We now proceed to obtain the Green's function (\ref{eq:pm2}). Let us first consider the case in which $V$ vanishes linearly in radius at all turning points. This is the generic situation in which we are not at a boundary between two of the three regions above. The matching computation is standard and we have relegated the details to appendix \ref{sec:matchingeasy}. The result for the 
retarded Green's function at leading order in the WKB expansion is, in cases I and III,
\be\label{eq:13text}
G^R(\w,k) \propto \frac{i}{2} \lim_{r \to 0} r^{- 2 m L} \exp\left\{- 2 \g \int^{r_{1/3}}_r dr \sqrt{V} \right\}  \,, \qquad \text{(cases III and I)} \,.
\ee
This expression is of course precisely what we should expect. The Green's function is pure imaginary and given by the probability of the particle tunneling through to the interior oscillatory region and into the horizon. This probability is exponentially small.

In case II the fermion propagates through an intermediate oscillatory region. The Green's function is found to be
\be\label{eq:twotext}
G^R(\w,k) \propto \frac{i \, {\mathcal G}}{2} \lim_{r \to 0} r^{- 2 m L} \exp\left\{- 2 \g \int^{r_1}_r dr \sqrt{V} \right\}  \,, \qquad \text{(case II)}
\ee
where
\be\label{eq:curlyG}
{\mathcal G} = \frac{\cosh(X + i Y) + \sinh(X - i Y)}{\cosh(X - i Y) - \sinh(X + i Y)} \,,
\ee
where here
\be\label{eq:XY}
X = \g \int_{r_2}^{r_3} dr \sqrt{V} + \log 2 \,, \qquad Y = \g \int_{r_1}^{r_2} dr \sqrt{-V} + \frac{\pi}{2} \,.
\ee

The Green's function in case II has poles when $i e^{- 2 X} = \tan Y$. Note that both $X$ and $Y$ are large positive numbers when $\w$ is real. Including the exponentially small imaginary part of the pole to leading order, the poles are at
\be\label{eq:poles}
Y = \pi n + i e^{- 2 X} \,.
\ee
Here $n$ is an integer. This formula has an immediate interpretation. For a given momentum $k < k_F^\star$ it is to be solved for the `Bohr-Sommerfeld' frequencies $\w_n$. These frequencies are closely spaced and have a real part much greater than their imaginary part. These resonant frequencies correspond to exciting a fermion that is making up the electron star background. The small imaginary part gives the probability for the excitation to tunnel through the barrier and fall into the event horizon. From a field theory perspective this is the probability of dissipating into the IR quantum critical modes. In the background electron star itself, this process is prevented through the electrostatic repulsion between electrons as well as the degeneracy pressure.

Let us emphasize that there are two distinct exponential suppressions here. The first is common to all cases and is due to the fact that the Green's function is a boundary Green's function, that the potential is positive at the boundary, and that therefore the fermion needs to tunnel into the spacetime. This suppression is an overall factor, yet it will play an important r\^ole shortly as it means that some quasiparticle poles have a residue that is exponentially smaller than others, depending on how far the fermion has to tunnel. The second exponential suppression is the imaginary part of the poles appearing in (\ref{eq:poles}). This is the statement that the bound states in the star are very stable as they are separated from the horizon by a potential barrier.

The next step is to translate the above results into explicit statements about the frequency and momentum dependence of the retarded Green's function. We are interested in low frequencies but arbitrary momentum. In this regime we have noted that there are three distinct cases
\be
\begin{array}{ccccc}
k > k_F^\star & \leftrightarrow & \text{Case I} & \leftrightarrow & \text{high momentum} \,, \\
k_F^\star > k \gtrsim \w^{1/z} & \leftrightarrow & \text{Case II} & \leftrightarrow & \text{intermediate momentum} \,, \\
\w^{1/z}\gtrsim k  & \leftrightarrow & \text{Case III} & \leftrightarrow & \text{low momentum} \,. \\
\end{array}
\ee
See also figure \ref{fig:wkplane} above.

Case I is familiar in essence from Fermi liquids. At zero temperature and at momenta much greater than the Fermi momentum, there are no low energy particle or hole excitations. Therefore the Green's function should not have an interesting pole structure in the frequency dependence. From the expression for the Green's function (\ref{eq:13text}) we can obtain the leading low frequency dependence. Outside of the near horizon Lifshitz region, all quantities are analytic as $\w \to 0$. However, the turning point $r_3$ is inside the Lifshitz region. As $\w \to 0$, the turning point necessarily has $r \to \infty$. In this regime we may approximate (\ref{eq:nearE}) by
\be
V = g_\infty \left(\hat k^2 - r^{2z-2} \hat \w^2 \right) \,.
\ee
The turning point is therefore
\be
r_3 = \left(\frac{k}{\w} \right)^\frac{1}{z-1} \,.
\ee
Using the above two expressions for the integral in (\ref{eq:13text}) we can easily extract the singular frequency dependence of the Green's function. 
\be\label{eq:aa}
G_I^R(\w,k) \propto i \exp\left\{ - 2 \eta \, \g \left(\frac{k^z}{\w} \right)^\frac{1}{z-1}  \right\} \,, \ee
where we are working up to an overall momentum-dependent constant and the constant in the exponent is
\be\label{eq:bb}
\eta = \frac{\sqrt{\pi g_\infty}}{2 z} \frac{\Gamma\left(\frac{1}{2(z-1)} \right)}{\Gamma\left(\frac{z}{2(z-1)} \right)} \,.
\ee
The expressions (\ref{eq:aa}) and (\ref{eq:bb}) are our final answer for the retarded Green's function in case I. The same formulae were derived, for a slightly different question, in \cite{Faulkner:2010tq}. The spectral density has no features and is exponentially small at low frequencies but has a branch cut at $\w=0$ due to the possibility of dissipating (inefficiently) into the quantum critical Lifshitz modes.

Case II is both similar and distinct from Fermi liquids. In a Fermi liquid there is a unique momentum $k_F$ close to which there are low energy particle and hole excitations. This leads to a pole in the Green's function at $\w=0$ and $k=k_F$. In the non-Fermi liquids dual to electron stars, the Fermi surface is `smeared' \cite{Hartnoll:2010xj} into a near-continuum of Fermi surfaces. These correspond to bulk Fermi surfaces at differing radii in the electron star. This is a large $N$ holographic manifestation of the smearing of spectral density due to strong interactions with critical bosonic modes. As we might have anticipated, we see that the Fermi liquid pole at a single momentum and energy is replaced by a sequence of poles closely spaced in energy for any given momentum (\ref{eq:poles}). This is because any given momentum (not too far) below $k_F^\star$ is the Fermi momentum at two radii in the bulk. There are therefore gapless excitations at these radii as well as arbitrarily light excitations at nearby radii.

Here we should emphasize that the above results show (unsurprisingly) that a WKB treatment is powerful enough to identify a large number of closely spaced yet discrete poles, rather than a continuum. We will have to do some work below in order to re-establish contact with the picture of a smeared Fermi surface.

As we wish only to capture nonanalytic dependences on frequency and momentum, we can zoom in on the poles (\ref{eq:poles}) of the retarded Green's function (\ref{eq:two}). Thus we can write the prefactor (\ref{eq:curlyG}) as
\be\label{eq:polesA}
{\mathcal G} = \sum_n \frac{i}{Y(\w,k) - \pi n + i e^{- 2 X(\w,k)}} \,.
\ee
Recall that $Y$ includes a term plus $\half \pi$ in (\ref{eq:XY}).
The real part of the denominator vanishes at momenta $k_F^{(n)}$ where
\be\label{eq:flpoles}
Y(0,k_F^{(n)}) = \pi n \,.
\ee
These are the many Fermi momenta.
Recall that $Y$ is large in the WKB limit and therefore these poles are reliable for large values of $n$. The poles are therefore closely spaced in momentum. We can now expand $X$ and $Y$ and the overall integral in (\ref{eq:twotext}) about $\w=0$ and $k = k_F^{(n)}$ to obtain an expression for the Green's function that captures all of the low frequency singular behaviour:
\be\label{eq:twofinal}
G_{II}^R(\w,k) \propto \sum_{0 \, < \, k_F^{(n)} < \, k_F^\star} \frac{c_n \g^{-1} e^{- 2 \g a_n}}{\; \w + v_n (k - k_F^{(n)}) + i c_n \g^{-1} \exp\left\{- 2 \g \left[ \eta \left(k_F^{(n)\,z}/\w \right)^\frac{1}{z-1} + b_n \right]\right\}} \,.
\ee
Here $a_n, b_n, c_n, v_n$ are order one (in $\gamma$) positive numbers that can be obtained from the integrals appearing in (\ref{eq:twotext}) at $\w=0$ and $k=k_F^{(n)}$. In particular
\be
v_n = \frac{\pa_k Y(0,k_F^{(n)})}{\pa_\w Y(0,k_F^{(n)})} \,, \qquad c_n = \frac{\g}{\pa_\w Y(0,k_F^{(n)})} \,,
\ee
and
\be
a_n = \int_0^{r_1} \left(\sqrt{V(0,k_F^{(n)})} - \frac{\hat m}{r} \right) dr + \hat m \log r_1 \,.
\ee
Here we added and subtracted a term $\hat m/r$ in the integrand so that we could take the integral all the way to the boundary at $r=0$. The constant $\eta$ in (\ref{eq:twofinal}) is again given by (\ref{eq:bb}). The answer agrees precisely with the `semi-holographic' result \cite{Faulkner:2010tq} with a Lifshitz IR geometry, as it must. The singular frequency dependence of the imaginary part of the denominator follows from the same argument that was outlined for case I above. The numerator has no singular frequency dependence because the turning point for the overall integral in (\ref{eq:twotext}), namely $r_1$, is in the far region where the zero frequency limit may be taken smoothly.

Equation (\ref{eq:twofinal}) is our final expression for the retarded Green's function in case II. It describes a large number of Fermi surfaces making up the star, with Fermi momenta bounded between zero and the extremal Fermi momentum $k_F^\star$ of (\ref{eq:KF}). These fermions are exponentially long lived at low frequencies. The spectral density will therefore be well approximated by a sum of delta functions. Note however that the residue of the poles in (\ref{eq:twofinal}) depends exponentially on $\g$ times a function of the Fermi momenta.

\begin{figure}[h]
\begin{center}
\includegraphics[height=140pt]{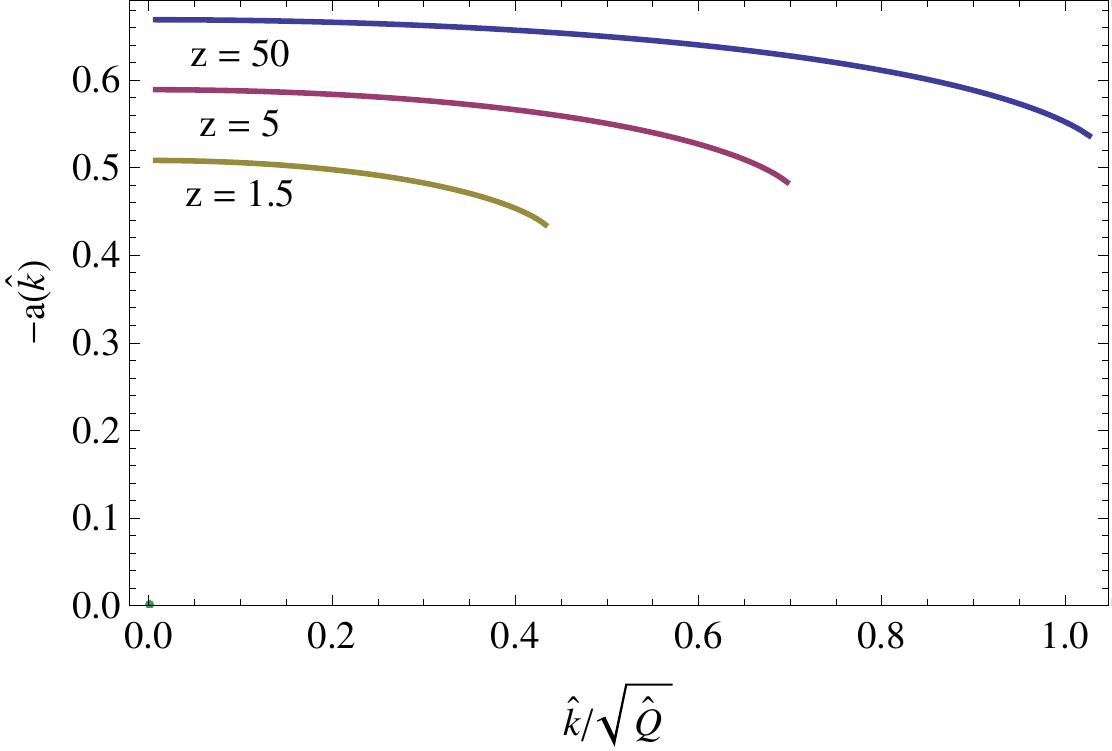}\hspace{0.2cm}\includegraphics[height=140pt]{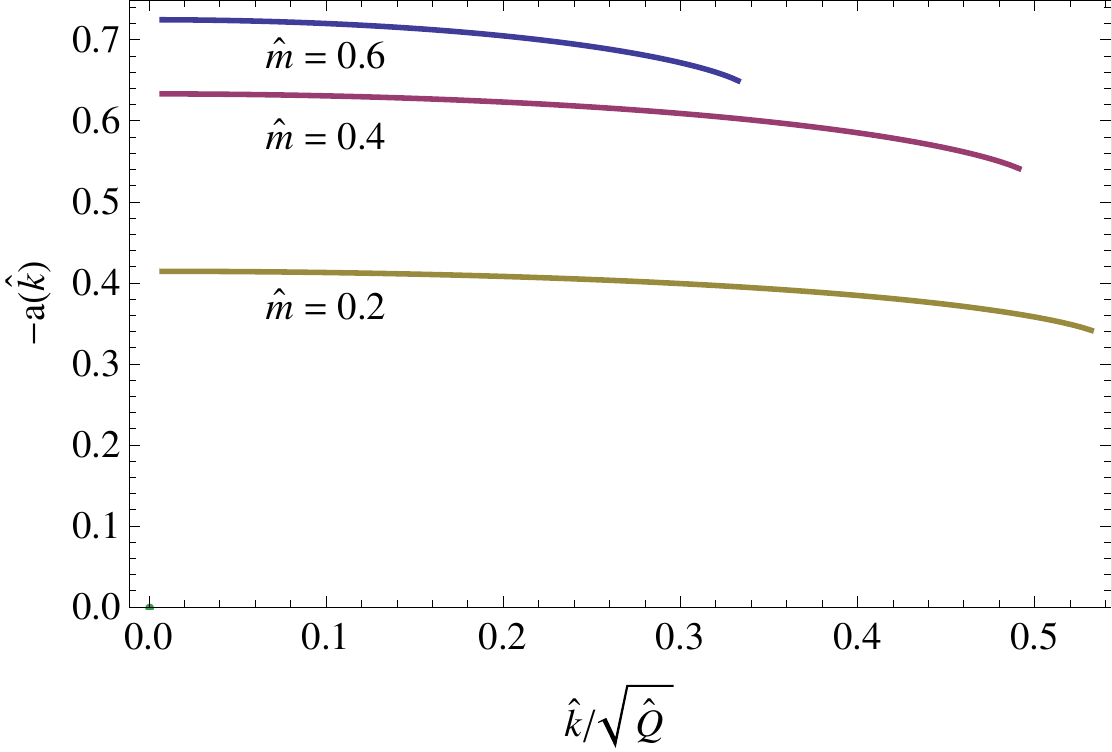}\caption{Exponent of the residue of the Fermi surface poles as a function of Fermi momentum (in units of the charge density). Left hand plot has $\hat m = 0.3$ for all curves while the right hand plot has $z=2$ for all curves. The plots show that Fermi surfaces near the maximal Fermi momentum have exponentially small spectral weight compared the other poles. \label{fig:residues}}
\end{center}
\end{figure}

The plots in figure \ref{fig:residues} show the exponent $-a_n$ of the Fermi surface spectral weight, $e^{- 2 \g a_n}$ in (\ref{eq:twofinal}), as a function of the Fermi momenta $k_F^{(n)}$ for several values of the electron star parameters. We see that the Fermi surfaces at small momenta, near the top of the potential, have exponentially larger spectral weight than those near $k_F^\star$, at the bottom of the potential.
We can understand this intuitively from the plot of the effective Schr\"odinger potential in figure \ref{fig:cases}; the fermion has further to tunnel from the boundary through to the bottom of the potential compared to high up in the potential. On the other hand, states near the bottom of the potential are exponentially more stable than those at smaller $k_F^{(n)}$ according to (\ref{eq:twofinal}). This is because they have further to tunnel to reach the interior Lifshitz geometry. Low energy observables can depend on both the lifetime and the spectral weight of the Fermi surface excitations. Depending on which exponential dominates for a given observable (i.e. lifetime or weight), the dominant fermion contribution will come from very different excitations of the electron star.

Case III is distinct from Fermi liquids. At sufficiently small momenta even the `smeared' Fermi surface is ceasing to exist. The appearance of the IR quantum critical dispersion relation $\w \sim k^z$ suggests an interpretation for the physics at these momenta. Namely, when $\w \gtrsim k^z$ the fermionic excitations are able to dissipate efficiently into on shell quantum critical modes and therefore lose their quasiparticle residue.
In this case, from equation (\ref{eq:13text}) the Green's function has only analytic $\w$ dependence. This follows from the fact that $r_1$ is not in the near horizon region, i.e. that, unlike case I, the fermions do not tunnel directly into the Lifshitz sector, and therefore the integrand may be expanded in analytic powers of small $\w$ over the whole integration range. Our final expression for the low frequency Green's function in this regime is simply
\be\label{eq:Giii}
G_{III}^R(\w,k) \propto i e^{- 2 \g \, c} \,.
\ee
Here we suppress analytic momentum and frequency dependence, to leading order the exponent is simply a constant $c$. We should recall however that the boundary of case III is itself defined via a non-analytic relation (\ref{eq:three}) between frequency and momentum: $\w \gtrsim k^z$.

The above analysis breaks down at frequencies and momenta for which turning points appear or disappear. These are the boundaries between cases I and II and between cases II and III. We will shortly see how interesting frequency dependences emerge at these boundaries. Before turning to these points however, we can note an interesting way in which the Luttinger count works out from the singularity structure of the fermion Green's function.

\section{The Luttinger count}

The Luttinger theorem relates the total charge density of a Fermi liquid to the sum of the volumes of Fermi surfaces \cite{oshikawa}.
For a spin half fermion in 2+1 dimensions with Fermi surfaces with momentum space volumes $V_n$
\be
Q = \sum_n \frac{2}{(2\pi)^2} V_n \,.
\ee
This `Luttinger count', in which the UV charge density is recovered in terms of the IR Fermi surfaces can be violated if the microscopic fermions in the system are coupled to gauge fields. A recent discussion of this fact can be found in \cite{Huijse:2011hp}. As well as Fermi liquid-like states, in which all the charge is visible in gauge-invariant Fermi surfaces, it is also possible to have `Fractionalized Fermi liquids' \cite{fraction} in which the sum of gauge-invariant Fermi volumes is less than the total charge. It is also emphasized in \cite{Huijse:2011hp} that the best established examples of gauge-gravity duality have the structure necessary to admit Fermi-liquid like (i.e. satisfying the Luttinger theorem) as well as fractionalised Fermi liquid phases. Now that we have characterised the singularity structure of the fermion correlators of the fermions making up the star, we can ask whether the Luttinger count holds true. We can thereby determine what type of phase we are in.

In \cite{Hartnoll:2010xj}, an argument was given indicating that the Luttinger count did work out for electron stars.\footnote{This statement should not be confused with an additional statement made in \cite{Hartnoll:2010xj}, which is that the Fermi surface area as extracted by quantum oscillations would not satisfy a Luttinger count.} All of the charge is carried by fermions in the bulk. These fermions are locally Fermi liquids and therefore at each radius satisfy a Luttinger theorem in the bulk. Thus the total charge will be given by the sum over bulk Fermi volumes. This argument is not entirely satisfactory, however, because the local bulk Fermi volumes are three dimensional while the boundary Fermi surfaces are two dimensional. With the boundary field theory fermion Green's functions at hand we are now in a position to make a more precise statement.

As shown in \cite{Hartnoll:2010gu}, the charge density of the field theory is given by integrating over all the fermions in the star
\be\label{eq:Qold}
Q = \frac{1}{3 \pi^2} \int^\infty_{r_s} dr \, r \sqrt{g} \, k_F^{3}(r) \,. 
\ee
The local Fermi momentum $k_F(r)$ was defined in (\ref{eq:KFr}).
Note that there are no hats over any of these quantities. We have also used the fact that $\beta$ defined in \cite{Hartnoll:2010gu} is equal to $\pi^{-2}$ for spin half fermions. This expression is compatible with a bulk (three spatial dimensional) Luttinger count at each point because we could write it as
\be
Q =  \int^\infty_{r_s} dr \, r \sqrt{g} \frac{2}{(2 \pi)^3} V(r) \,,
\ee
where here $V(r) = \frac{4 \pi}{3} k_F^3(r)$ is the Fermi surface volume at the radius $r$.

From a 2+1 dimensional field theory perspective we need to evaluate
\be\label{eq:nsum}
\sum_n \frac{2}{(2 \pi)^2} V_n = \sum_n \frac{1}{2 \pi} k^{(n) \, 2}_F\,,
\ee
where $k^{(n)}_F$ are the Fermi momenta appearing in our Green's function. They are defined by poles of the Green's function at $\w = 0$. From (\ref{eq:XY}) and (\ref{eq:flpoles}), and more generally the discussion of the previous section, we have
\be\label{eq:wkbn}
\int_{r_1}^{r_2} dr \, r \sqrt{g} \sqrt{k_F^2(r) - k^{(n)\, 2}_F} = \pi \left(n + \half \right) \,.
\ee
There is no overall $\g$ on the right hand side because there are no hats on the momenta in the square root, cf. (\ref{eq:wkhat}).
The order one contribution to the Luttinger sum comes from highly excited states in our WKB limit and so we can approximate the sum over $n$ in (\ref{eq:nsum}) by an integral. Letting
$k^{(n)\, 2}_F = E(n)$ we have
\bea
\sum_n \frac{2}{(2 \pi)^2} V_n & = & \frac{1}{2\pi} \int dn E(n) = \frac{1}{2\pi} \int_0^{k_F^{\star \, 2}} dE E \frac{dn}{dE} \\
& = & \frac{1}{(2\pi)^2} \int_0^{k_F^{\star \, 2}}  dE E \int_{r_1}^{r_2} dr \, \frac{r \sqrt{g}}{\sqrt{k^2_F(r) - E}} \\
& = & \frac{1}{(2\pi)^2} \int_{r_s}^{\infty} dr \, r \sqrt{g} \int_{0}^{k_F^2(r)} dE \frac{E}{\sqrt{k^2_F(r) - E}} \\
& = & \frac{1}{3 \pi^2} \int_{r_s}^{\infty} dr \, r \sqrt{g} \, k^3_F(r) \\
& = & Q \,.
\eea
In the first line, $k_F^\star$ is the maximum of $k_F(r)$ as defined in (\ref{eq:KF}).
To get from the first to the second line we differentiated the expression (\ref{eq:wkbn}) with respect to $n$. From the second to the third line we exchanged the orders of integration, being careful with limits. We then perform the integral over $E$ in the third line. The final result then follows from (\ref{eq:Qold}).

The computation we have just performed confirms that the Luttinger count holds for electron stars: the sum of the areas of the gauge invariant field theory Fermi surfaces adds up to the total charge density. There is a something slightly magical almost about the way that the square root appearing in the standard WKB formula (\ref{eq:wkbn}) acts as a projection from the spherical bulk Fermi surfaces to the circular boundary theory Fermi surfaces in such a way that the factors of $\pi$ work out.

We end this section by emphasizing that while the Luttinger count works out, the dual state of matter described by the electron star is not a Fermi liquid. In particular, both the spacetime geometry and the fermion Green's function indicate that there is a critical Lifshitz sector with gapless degrees of freedom interacting with the fermionic excitations at arbitrarily low energies and momenta.

\section{The boundary fermion Green's function:  $k \sim k_F^\star$}
\label{sec:kftext}

This section computes the fermion Green's function at low frequencies and at momenta on the boundary between cases I and II above. Recall that this corresponds to momenta close to the extremal Fermi momentum $k_F^\star$. This is the largest of many Fermi momenta of the electron star. It was shown in \cite{Hartnoll:2010xj} that this momentum uniquely determined the period of de Hass-van Alphen quantum oscillations of the system. We expect that the Fermi momentum poles will coalesce into a branch cut in the smearing limit. We may expect that one endpoint of the cut will be precisely at $k \sim k_F^\star$ and therefore we need to characterise these poles carefully.

The WKB approximation used in the previous section fails for momenta sufficiently close to $k_F^\star$. The reason for this is that once the turning point becomes sufficiently close to the minimum of the potential one cannot treat the potential as locally linear. Specifically, near the minimum we can write the potential as
\be\label{eq:AB}
V = - A + \frac{B^2}{4} (r - r_o)^2 \,.
\ee
The radius $r_o$ is slightly shifted from $r_\star$ as we show in appendix \ref{sec:kf}.
Our discussion will be more transparent if we rescale variables and introduce
\be\label{eq:rescale}
y = \sqrt{\g B} (r - r_o) \,, \qquad \e = \g A/B \,.
\ee
The Schr\"odinger problem then becomes
\be\label{eq:Vmin}
- \ddot \Phi + \left(-\e + \frac{y^2}{4} \right) \Phi = 0 \,. 
\ee
Here dots denote derivatives with respect to $y$. Let us linearise near the turning point and determine whether WKB is valid. 
Set $y =  2 \sqrt{\e} + \d y$. To expand and obtain a linear potential we require $\d y \ll \sqrt{\e}$. Expanding, the equation becomes $- \ddot \Phi + \sqrt{\e}\d y \Phi = 0$. Now plugging into the conditions for WKB to hold (\ref{eq:wkbconditions}), and recalling to convert the derivatives with respect to $r$ into derivatives with respect to $y$, we see that WKB requires $\e^{-1/6} \ll \d y$. From the two inequalities we have just obtained, it follows that we can only consistently perform a matching to WKB across a linear matching point close to a minimum of the potential when
\be\label{eq:linearyes}
1 \ll \e \,. \qquad \qquad \text{(for linear matching)} \,.
\ee
This condition will not hold for $k$ sufficiently close to $k_F^\star$ where $\e$ tends to zero.

When (\ref{eq:linearyes}) does not hold we will instead solve the Schr\"odinger equation in the full quadratic region where (\ref{eq:Vmin}) holds and match this solution directly onto WKB on each side. The matching solution is now given by
Parabolic Cylinder $D$ functions (this is just a fancy name for a combination of Confluent Hypergeometric functions), and not the Airy functions of the simpler linear matching. In order for this matching to be valid, we need to be able to neglect third and higher order corrects to the potential (\ref{eq:AB}). Generically, recall the rescaling (\ref{eq:rescale}), this will require $y \ll \sqrt{\g}$. Furthermore, in order to be able to match, we need the quadratic form (\ref{eq:Vmin}) of the potential to hold simultaneously with the WKB conditions (\ref{eq:wkbconditions}). A sufficient condition for satisfying the WKB conditions together with $y \ll \sqrt{\g}$ is that
\be\label{eq:quadraticyes}
\e \ll \g \,. \qquad \qquad \text{(for quadratic matching)} \,.
\ee
Fortunately we see that (\ref{eq:linearyes}) and (\ref{eq:quadraticyes}) overlap in our large $\g$ limit and therefore we can cover the whole range of momenta by using one or the other method.

To implement the quadratic matching we need to expand the potential (\ref{eq:potential}) about its minimum. At $\w=0$ and $k = k_F^\star$, see equations (\ref{eq:KF}) and (\ref{eq:VKF}) above, the minimum is at $r = r_\star$ and $V(r_\star)=0$. We are interested in the potential at nearby values of frequency and momentum, $\hat \w = \d \hat \w$ and $\hat k = \hat k_F^\star + \d \hat k$. In appendix \ref{sec:kf} we show that the potential takes the form (\ref{eq:AB}) with
\be\label{eq:AB2}
A = 2 g(r_\star)\left(\frac{h(r_\star)}{f(r_\star)} \d \hat \w - \hat k_F^\star r_\star^2 \d \hat k \right) \,, \qquad \text{and} \qquad B = 2 r_\star \sqrt{- g(r_\star) \hat k_F^\star \hat k_F''(r_\star)} \,.
\ee
Recall that $r_\star$ is defined via (\ref{eq:KF}) and $\hat k_F(r)$ is given by (\ref{eq:KFr}).
From here we can obtain $\e$ via (\ref{eq:rescale}). Before presenting the full Green's function, note that the insofar as we are interested in the low lying normal modes and neglect exponentially small tunneling probabilities, we can simply assume that the wavefunction is normalisable in the quadratic region. This corresponds to the familiar Hermite functions and hence we obtain a spectrum of fermionic excitations
\be\label{eq:en}
\e \sim n+ \frac{1}{2} \,, \qquad \qquad (n>0) \,.
\ee
For $1 \ll n$ these are the poles we obtained in our previous analysis (\ref{eq:polesA}). We have just shown that the approach of this current section allows us to go to lower poles $n \ll \g$. At larger values of $n$ the states probe the deviation of the potential away from the purely quadratic form near the minimum.

If we set $\delta \hat \omega=0$ we can think of equation (\ref{eq:en}) as determining the Fermi momenta of each species of particle in our star. Using the above few equations:
\be\label{eq:below}
\hat k_F^{(n)} = \hat k_F^\star - \left(\frac{-\hat k_F''(r_\star)}{g(r_\star) r_\star^2 \hat k_F^\star}\right)^{\frac{1}{2}} \left(\frac{n}{\g}+\frac{1}{2\g}\right) \,.
\ee
The $+\frac{1}{2\g}$ term here indicates that the extremal Fermi momentum $\hat k_F^\star$ receives a quantum correction. As explained below equation (\ref{eq:schro}) above, there are additional corrections at this order that we are not considering, so the quantum shift cannot be considered too literally. In the large $\gamma$ limit the spacing between states is suppressed, leading to a pseudo continuum distribution of poles. These poles merge into the poles of (\ref{eq:polesA}), with the two descriptions overlapping for $1 \ll n \ll \g$. We shall verify the agreement shortly.

Similarly if we set $\delta \hat k = 0$ we obtain a spectrum of frequencies $\delta \hat \omega \propto \frac{n}{\g} + \frac{1}{2 \g}$.
These poles at positive frequencies corresponding to the cost of creating particles of momentum $\hat k_F^\star$ for the different species of particles at different radii in the star. At $\hat k_F^\star$ all Fermi seas of all species are empty, so it is not possible to destroy a particle of this momentum. That is why there are no negative frequency poles.

Our discussion so far has neglected the exponentially small imaginary part of the poles. We will now compute the full Green's function for $k \sim k_F^\star$, including the real and imaginary part of the poles and their residues. We have relegated the details to appendix \ref{sec:kf}. Here we simply quote the result for the retarded Green's function
\be\label{eq:green1text}
G^R_{\text{I-II bdy.}}(\w,k) \propto e^{-2\gamma I_B - 2\e T_B} e^{i \e \pi}  e^{- i \frac{\pi}{2}} \frac{1 -  \frac{1}{2} \frac{\sqrt{2 \pi}}{\Gamma\left(\frac{1}{2}+\e \right)}e^{-2\gamma I_S - 2\e T_S}}{\frac{\sqrt{2\pi}}{\Gamma\left(\frac{1}{2}-\e \right)} - \frac{1}{2}e^{i \e \pi}  e^{-2\gamma I_S - 2\e T_S} } \,.
\ee
Recall $\e$ was introduced in (\ref{eq:rescale}) in terms of $A$ and $B$ which are given in (\ref{eq:AB2}) and which characterise the minimum of the potential according to (\ref{eq:AB}). The overall exponential terms characterising the tunneling into the spacetime from the boundary are
\begin{eqnarray}
I_B &=&  \int_{0}^{r_o} \left( \sqrt{\Delta V} - \frac{\hat m}{r} \right) \,dr + \hat m \log r_o \,, \\
T_B &=& \int_{0}^{r_o}\left( - \frac{B}{2 \sqrt{\Delta V}} + \frac{1}{ |r-r_o|} \right)\,dr -  \log{ \sqrt{\gamma B}  r_o} \,.
\eea
Here we set $V = - A + \Delta V$. The first line includes terms that regulate the integral near the boundary $r \to 0$. The exponential terms that measure the tunneling through the star to the horizon, giving an imaginary part to our poles, have exponents
\begin{eqnarray}
I_S &=&  \int^{r_3}_{r_o} \sqrt{\Delta V} \,dr  \,, \\
T_S &=& \int^{r_3}_{r_o}\left( - \frac{B}{2 \sqrt{\Delta V}} + \frac{1}{ |r-r_o|} \right) \,dr -  \log{ \sqrt{\gamma B}  |r_3-r_o|} \,.
\end{eqnarray}
Here recall that $r_3$ is the (linear in this case) turning point closest to the horizon.

If we ignore the exponentially small imaginary parts in the Green's function (\ref{eq:green1text}) it is clear that we recover the poles at $\e = \frac{1}{2} + n$ described above. Furthermore, we can `smear' over the poles by looking at the Green's function in the regime $1 \ll \e \ll \g$. 
This is the regime of overlap of the quadratic and linear WKB matching, so we did not have to do a quadratic matching to access this regime. Here the poles should merge and form a branch cut. Using Stirling's expansion for the gamma functions in (\ref{eq:green1text}) and noting that $B$ is order one in $\g$, one obtains the leading order result
\bea
G^R_{\text{I-II bdy.}}(\w,k) & \sim & i e^{-2\gamma I_B + \e \log (\g/\e)} \,, \qquad \qquad \quad (1 \ll \e \ll \g) \,. \\
 & \sim & - i e^{-2\gamma I_B- \# \g (\hat \w + v_\star (\hat k_F^\star - \hat k)) \log (\hat \w + v_\star (\hat k_F^\star - \hat k))} \,. \label{eq:smear}
\eea
Here $\#$ and $v_\star$ do not depend on the WKB parameter, the frequency or momentum. Thus we see that smearing over the poles does lead to a branch cut emanating from the extremal Fermi surface at
\be
\hat \w \sim v_\star (\hat k_F^\star - \hat k) \,.
\ee
The singularity at this extremal momentum is weak. The spectral weight in the branch cut increases exponentially as we move to lower momentum $\hat k$, away from the branch point. This is consistent, as it had to be, with our discussion around figure (\ref{fig:residues}) above; the residue of the Fermi surface poles close to $k_F^\star$ are exponentially small. Nonetheless, the extremal Fermi momentum does produce a nonanalyticity in the smeared Green's function (\ref{eq:smear}). This can then lead to important signatures such as quantum oscillations \cite{Hartnoll:2010xj} that pick out momentum space nonalyticities and are independent of the spectral weight. The disjunction between quantum oscillations and spectral weight is an interesting feature of electron stars.

\section{The boundary fermion Green's function:  $k \sim \w^{1/z}$}
\label{sec:wktext}

This section computes the fermion Green's function at the boundary of cases II and III above. Recall from equation (\ref{eq:three}) that this corresponds to small frequencies and momenta close to a dispersion relation $k \sim \w^{1/z}$. For momenta below the dispersion relation, the fermion states become highly unstable to dissipation into the quantum critical sector. In the Schr\"odinger equation this translates into the fact that we will be considering states close to the top of the effective potential. In contrast, the previous section considered the most stable states, at the bottom of the potential. Our computation is similar to that of the previous section, except that now we must match the WKB region onto an inverted harmonic oscillator potential.
A further potential complication is that while we are taking $\gamma$ large, as usual, we also need to take $\omega$ and $k$ very small as they scale towards the critical dispersion relation.

We will be expanding around the top of the barrier, of for instance figure \ref{fig:cases}, that prevents fermion states from falling into the black hole. Near the maximum the potential takes the form
\be\label{eq:ABmax}
V = - A - \frac{B^2}{4} (r-r_o)^2 \,.
\ee
The difference from (\ref{eq:AB}) of the previous section is the opposite sign of the quadratic term. Of course $A,B,r_o$ take different values as we shall see shortly. Again rescaling as in (\ref{eq:rescale}) we obtain the Schr\"odinger problem near the maximum
\be
- \ddot \Phi + \left(-\e - \frac{y^2}{4} \right) \Phi = 0 \,. 
\ee
Also as previously, linear matching close to the maximum of the potential is legitimate when $1 \ll \e$ while quadratic matching, the topic of this section, requires $\e \ll \g$.

We noted above that this maximum necessarily occurs in the near horizon Lifshitz region of the spacetime. The potential we are expanding about the maximum is therefore (\ref{eq:nearE}). The maximum is seen to occur at
\be\label{eq:xo}
x_o \equiv r_o^z \hat \w = \frac{1}{2(z-1)}\left(h_\infty (2-z) \pm \sqrt{h_\infty^2 z^2 - 4 \hat m^2 (z-1)} \right) \,.
\ee
where we should pick the plus sign to find a solution for positive $\hat \omega$. We saw above that case III never arises at negative $\hat \w$. We have introduced $x_o$ here; it is simply a number with no $\hat \w$ dependence. Expanding the potential about the maximum we can read off $A$ and $B$ in (\ref{eq:ABmax}):
\bea
A & = & g_\infty \left( \frac{(h_\infty + x_o)^2 - \hat m^2}{x_o^{2/z}} \hat \w^{2/z}  - \hat k^2 \right) \,, \label{eq:AAA} \\
B & = & \hat \w^{2/z} \frac{2 \sqrt{z g_\infty}}{x_o^{2/z}} \sqrt{h_\infty (z-2) x_o + 2 (z-1) x_o^2} \,. \label{eq:BBB}
\eea
Setting $A=0$ gives us the precise dispersion relation separating cases II and III. The condition for the validity of matching WKB across the quadratic potential is $A/B \ll 1$, which is now seen to require that $\hat k^2/\hat \w^{2/z}$ be close to the critical dispersion relation.\footnote{We should additionally worry whether $\hat \w$ and $\hat k$ small might invalidate the WKB analysis. The potentially dangerous assumption is that the quadratic term in the potential dominates all higher order terms consistently with the WKB conditions (\ref{eq:wkbconditions}). We can check that $\d r \ll r_o = (x_o/\hat \w)^{\frac{1}{z}}$ is sufficient for the quadratic term to dominate, while validity of WKB requires $\sqrt{\g B} \, \d r \gg 1$. Compatibility of these two inequalities implies $\sqrt{\g B} (x_o/\hat \w)^{\frac{1}{z}} \gg 1$. This condition is always true at large $\g$, the factors of $\hat \w$ drop out. We will consider below the case when $x_o$ becomes small.\label{foot:ha}}

In appendix \ref{sec:wk} we perform the matching across the quadratic region to obtain the Green's function. The result is
\be\label{eq:II3}
G^R_\text{II-III bdy.} \propto i e^{-2 \gamma I_B} \frac{\frac{e^{\ep \frac{\pi}{2}} \sqrt{2\pi}}{\G\left(\frac{1}{2} - i \ep\right)} e^{- 2 i \g I_W - 2 i \ep T_W} - 1}{\frac{e^{\ep \frac{\pi}{2}} \sqrt{2\pi}}{\G\left(\frac{1}{2} - i \ep\right)} e^{- 2 i \g I_W - 2 i \ep T_W} + 1}  \,.
\ee
Here $I_B$ again gives the amplitude to tunnel in from the boundary
\be\label{eq:iib}
I_B =  \int_{0}^{r_1} \left( \sqrt{V} - \frac{\hat m}{r} \right) \,dr + \hat m \log r_1 \,,
\ee
while the integrals characterising the periods of oscillation within the well are
\begin{eqnarray}
I_W &=& \int_{r_1}^{r_o} \sqrt{- \D V} \, dr \,, \label{eq:IW}\\
T_W &=& \int_{r_1}^{r_o} \left( \frac{B}{2 \sqrt{-\D V} } - \frac{1}{|r-r_o|} \right) \, dr + \log{\sqrt{\g B} |r_o-r_1|}\label{eq:TW} \,,
\end{eqnarray}
Recall from (\ref{eq:rescale}) that
\be\label{eq:epsolve}
\e = \g A/B \propto \g \left(\frac{(h_\infty+x_o)^2 - \hat m^2}{x_o^{2/z}} - \frac{\hat k^2}{\hat \w^{2/z}} \right) \,.
\ee

The previous section, in e.g. (\ref{eq:below}) and below, described how the exponentially stable fermionic excitations ended at the bottom of the potential in a harmonic oscillator fashion. We can now use the Green's function (\ref{eq:II3}) to see what happens to the poles near the maximum of the potential. In appendix \ref{sec:wk} we find the poles of the Green's function in the regime $1 \ll |\e| \ll \g$. We also derive simple expressions for $I_W$ and $T_W$ in (\ref{eq:TI}) that we use in the remainder. In this regime
the Green's function (\ref{eq:II3}) becomes
\be\label{eq:tan}
G^R_\text{II-III bdy.} \propto - e^{-2 \gamma I_B} \tan \left[ \frac{\ep}{2} \left(\log \frac{-\ep}{\g} - \frac{\sqrt{g_\infty(h_\infty^2- \hat m^2)}}{z} \frac{\g}{\e} \log \frac{1}{\hat \w} \right) \right] \,, \qquad  (1 \ll \ep \ll \g) \,.
\ee
Here we used the expressions in (\ref{eq:TI}) of appendix \ref{sec:wk} for $T_W$ and $I_W$.
Writing $\e$ in terms of real and imaginary parts
\be
\e = \e_1 + i \e_2 \,,
\ee
the poles are at
\bea
\e_1 & = & - \frac{2 \g \sqrt{g_\infty} \sqrt{h_\infty^2 -\hat m^2}}{z \log \g} \log \frac{1}{\hat \w} - \frac{\pi (2n + 1)}{\log \g}\,, \label{eq:spacingtext}\\
 \e_2 & = & \left\{ \begin{array}{ll}
0 & \quad \textrm{if  $\e_1<0$}\\ 
\displaystyle -\frac{\e_1 \pi}{\log{\gamma}}& \quad \textrm{if $\e_1>0$}
\end{array} \right. \,. \label{eq:ppp}
\eea
We see that, in this regime, while changing the value of (real) $\hat \omega$ shifts the poles around, they all lie in a line that tilts at the origin of the $\e$ complex plane. Ignoring the explicit $\hat \w$ dependence in (\ref{eq:spacingtext}), then setting $\e = \g A/B$ and solving for $\hat \w$ maps the line of poles with a finite imaginary part into a semicircle in the complex $\hat \omega^{2/z}$ plane. We see this semicircle vividly in the plots of section \ref{sec:smallstar} below. The explicit logarithm of $\hat \w$ only leads to mild violations of the semicircle form. A sketch of the poles described by (\ref{eq:ppp}) in the complex frequency plane is shown in figure \ref{fig:wkbpoles}. The plot is illustrative only, for (\ref{eq:ppp}) to be valid one should take a larger value of $\g$, which would see the second set of poles tilt only slightly into the lower half plane.

\begin{figure}[h]
\begin{center}
\includegraphics[width=330pt]{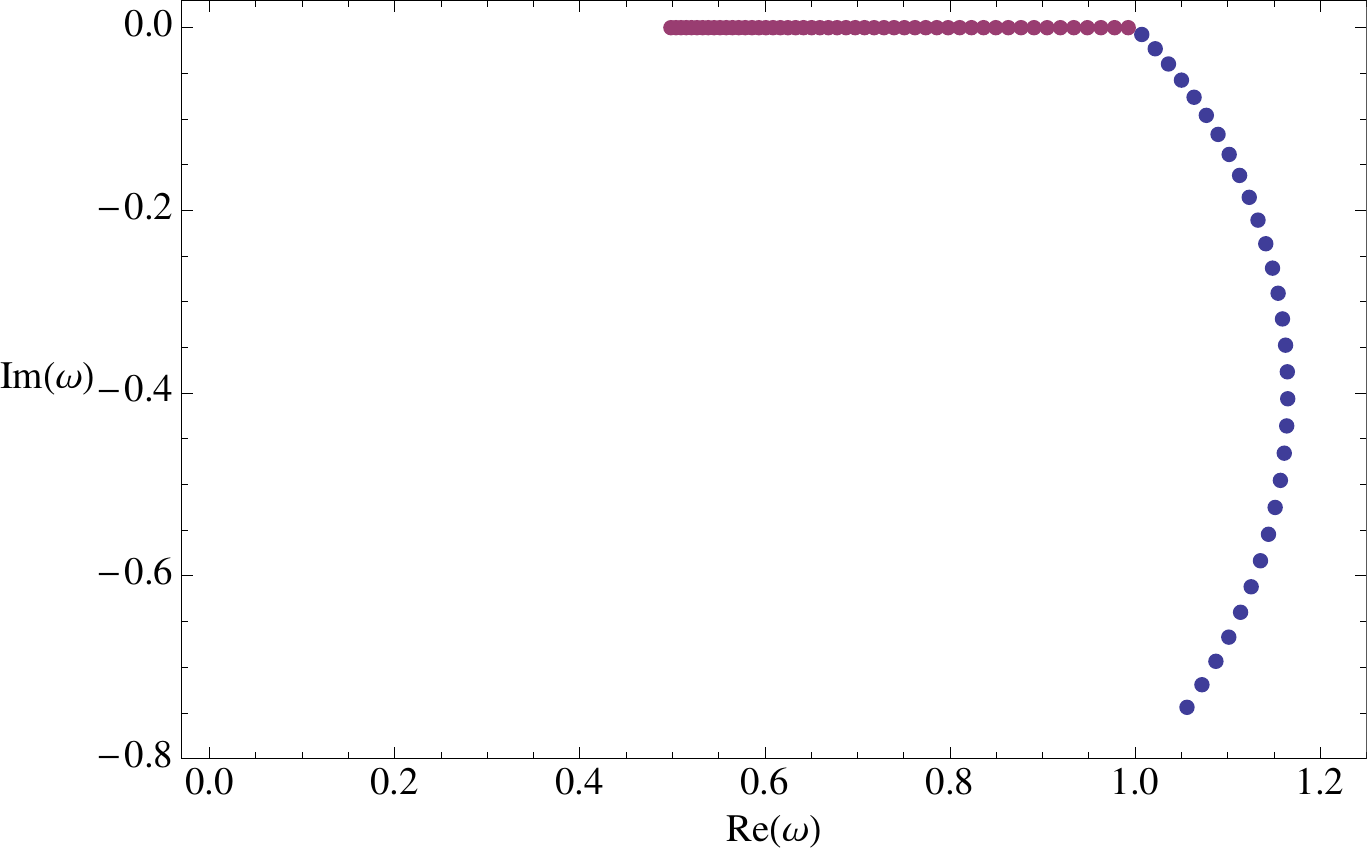}\caption{Typical distribution of the poles in the fermion Green's function close to $k \sim \w^{1/z}$, as described by (\ref{eq:ppp}). This plot has $\g = 100$ and $z=2$ and order one values for $\hat k, \hat m$, etc. We have not worried about carefully normalising the $\w$ axes. \label{fig:wkbpoles}}
\end{center}
\end{figure}

Unlike in the previous section, we have just seen that the limit $1 \ll |\e| \ll \g$ does not smear the Fermi surface poles into a branch cut, as the poles are still present and given by (\ref{eq:spacingtext}) and (\ref{eq:ppp}) and plotted in figure \ref{fig:wkbpoles}. Me must therefore work a little more to quantify how the Green's function changes at $\ep = 0$. The tangent appearing in the Green's function (\ref{eq:tan}) can be written as a sum over its poles
\be\label{eq:tanpoles}
\tan F(\ep) = - \sum_{n=-\infty}^\infty \frac{1}{F(\ep) + (n + \half) \pi} \,.
\ee
Now let us sit on (or parametrically close to) the real $\ep$ axis and ask whether we can approximate this sum by an integral. For $\ep < 0$ the function $F(\ep)$ in (\ref{eq:tanpoles}) is real to the order we are working and the tangent has poles on the real axis at (\ref{eq:spacingtext}). The spectral density (imaginary part of the Green's function) is thus a sum of delta functions and we should not replace the sum by an integral. Note that while the residue of the poles decreases as one moves away from $\e=0$, due to the overall $e^{-2 \gamma I_B}$ factor in (\ref{eq:tan}), it does not vary strongly over the spacing between poles and therefore one cannot keep only the first pole. For $\ep > 0$ however, $F(\ep)$ has an imaginary part of order $\ep$. We can therefore turn the sum over $n$ into an integral without coming close to any of the poles. Performing the resulting integral over $n$ by contour integration then gives the constant Green's function $G^R \propto i e^{-2 \gamma I_B}$ for $\ep > 0$.

It is clear from (\ref{eq:epsolve}) that $\ep$ changes sign at the dispersion relation
\be\label{eq:disper}
\hat \w(\hat k) = \hat k^{z} \frac{\left((h_\infty+x_o)^2 - \hat m^2 \right)^{z/2}}{x_o} \,.
\ee
So thus we find that for $\hat \w > \hat \w(\hat k)$ the Green's function $G^R \propto i e^{-2 \gamma I_B}$, in agreement with our previously obtained (\ref{eq:Giii}). Above the dispersion relation (\ref{eq:disper}) the fermion spectral density is featureless. Below it is characterised by closely spaced poles. The Green's function (\ref{eq:tan}) therefore interpolates between cases II and III as we should expect. We will also see this transition from poles to a constant explicitly in the following `small star limit' section.

\subsection*{Contribution to the optical conductivity}

While we are focussing on the fermion Green's function, it is of interest here to point
out a consequence of the existence of the scaling regime $\hat \w \sim \hat k^z$ for a more complicated observable: the optical (i.e. frequency dependent) conductivity. The contribution of the stable fermions to the optical conductivity was computed in \cite{Hartnoll:2010gu}. It was found that $\sigma(\Omega) \sim \Omega^2$ for small $\Omega$. Here we will estimate
the one loop contribution of the states around the critical dispersion relation to see if they can dominate the tree level result of \cite{Hartnoll:2010gu} at small frequencies. 

The strategy will be to isolate the critical contribution to the conductivity. Since the Green's function, given by (\ref{eq:II3}), has (almost) a scaling behavior, we will be able to estimate easily the contribution of this regime to the conductivity. We will follow closely the discussion in \cite{Faulkner:2010zz}, where one can also estimate the conductivity from scaling arguments.\footnote{One important difference between the present case and the one discussed in \cite{Faulkner:2010zz} is that, because we have Fermi surfaces at arbitrarily small momentum $k_F^{(n)}$, leading to the scaling form (\ref{eq:II3}), we can scale $k$ with $\Omega$, while in \cite{Faulkner:2010zz} the $k$ integral has to be dealt with before one can make a scaling argument.}

The starting point is the expression for the conductivity at one loop:
\be\label{cond1}
\sigma(\Omega) \sim \frac{1}{i \Omega} \int d\omega_1 \, d\omega_2 \, \frac{f(\omega_1) - f(\omega_2)}{\omega_1 - \omega_2 - \Omega - i \epsilon} \Lambda^2 \int d k \, k \,  {\rm Im \,}  G^R(\omega_1,k) {\rm Im \,} G^R(\omega_2,k) \,,
\ee
where $f(\omega)$ is the Fermi-Dirac distribution at zero temperature given by 1 at $\omega<0$ and 0 for $\omega>0$. $\Lambda$ is a radial integral over the interaction vertex and the fermion and gauge field wavefunctions  \cite{Faulkner:2010zz}. This integral extends, in principle, all the way from the boundary to the horizon of the holographic radial coordinate and it depends on the frequencies. Since we are only concerned with extracting an interesting contribution for $\sigma(\Omega)$  we can cut off the integral away from the horizon and consider a finite UV contribution where $\Lambda$ is analytic in $\Omega$. Furthermore, in the UV, it is clear that the wave functions have a smooth limit as $\Omega$ goes to zero and we can replace $\Lambda$ by a constant $ \Lambda_{UV}$, as in \cite{Faulkner:2010zz}.

The Green's function $G^R(\omega, k)$ in (\ref{eq:green}) is in fact just a function of $\ep$ and $I_W$ and $T_W$.  In appendix \ref{sec:wk} we show that $T_W$ is a $\gamma$ dependent constant, while $I_W$ behaves as $\log \omega^{\frac{1}{z}}$ in an appropriate limit. Let us then perform the following rescaling:
\be
\omega_1 = \Omega \, \tilde \omega_1\,,  \quad \omega_2 = \Omega \, \tilde \omega_2 \,, \quad k = \Omega^{\frac{1}{z}} \, \tilde k \,.
\ee
Under this scaling $\ep$ remains unchanged (\ref{eq:epsolve}). This means that the Green's function is of a scaling form $G(k^2/\omega^{\frac{2}{z}})$, up to log-periodic violations which we will neglect. The function $f(\omega)$ is also invariant under rescalings.  Then, we can write (\ref{cond1}) as:
\begin{eqnarray}\label{cond2}
\sigma(\Omega) &\sim&  \Omega^{2/z}  \int d\tilde\omega_1 \, d\tilde\omega_2 \, \frac{f(\tilde\omega_1) - f(\tilde\omega_2)}{\tilde\omega_1 - \tilde\omega_2 -1 - i \tilde\epsilon} \Lambda_{UV}^2 \int d \tilde k \, \tilde k \,  {\rm Im \,}  G\left(\tilde k^2/\tilde \omega_1^{2/z}\right)  {\rm Im \,}  G\left(\tilde k^2/\tilde \omega_2^{2/z}\right) \nonumber\\
&\sim& \Omega^{2/z} \, \gg  \, \Omega^2 \,.
\end{eqnarray}

The conclusion is that the contribution of critically scaling fermions to the optical conductivity dominates the contribution of stable Fermi surfaces in the small frequency limit.

\section{The `small star' limit}
\label{sec:smallstar}

In this section we consider a limit in which essentially the entire star is contained within the Lifshitz region. Because the potential is known analytically in this case, we are able to characterise aspects of the Green's function explicitly. This limit is achieved by taking $z$ and $\hat m$ such that the mass is only just below the critical mass above which the electron star ceases to exist (\ref{eq:i2}). Namely
\be\label{eq:lambda}
\fbox{
$\displaystyle \lambda^2 \equiv h_\infty^2 -\hat m^2 \ll 1 \,.$}
\ee
We expect that in this limit the electron star will be `small', in a sense which we will now make precise. This limit was considered in \cite{Hartnoll:2010gu}, as the large $\hat \beta$ limit.\footnote{The quantities are related by $\l^2 = \frac{6^{4/3} \hat m^{2/3} (1-\hat m^2)^{2/3}}{(2 \hat m^4 - 7 \hat m^2 + 6)^{2/3}} \frac{1}{\hat \b^{2/3}}$. It is clear that using $\lambda$ is more convenient for our purposes here.\label{foot:a}}

Let us think of keeping $\hat m$ fixed and expanding in $\l \ll 1$. Recalling from (\ref{eq:i1}) that $h_\infty^2 = \frac{z-1}{z}$, we have
\be\label{eq:zexp}
z = \frac{1}{1 - \hat m^2} + \frac{\l^2}{(1 - \hat m^2)^2} + \cdots \,, \qquad h_\infty = \hat m + \frac{\l^2}{2 \hat m} + \cdots \,.
\ee
The remaining constant characterizing the IR Lifshitz geometry is computed from the formulae in \cite{Hartnoll:2010gu} to be
\be
g_\infty =  \frac{6- 7 \hat m^2 + 2 \hat m^4}{6(1 - \hat m^2)^2} + \frac{(6- 7 \hat m^2 + 2 \hat m^4) (1 + 4 \hat m^2)}{12 \hat m^2 (1 - \hat m^2)^3} \l^2 + \cdots \,.
\ee

Also from \cite{Hartnoll:2010gu} we can recall that moving away from the IR Lifshitz region ($r \to \infty$) is characterised by an exponent $|\a|$ 
\be\label{eq:corrections}
f = \frac{1}{r^{2z}} \left(1 + f_1 \frac{1}{r^{|\a|}} + \cdots \right), \; g = \frac{g_\infty}{r^{2}} \left(1 + g_1 \frac{1}{r^{|\a|}} + \cdots \right), \; h = \frac{h_\infty}{r^{z}} \left(1 + h_1 \frac{1}{r^{|\a|}} + \cdots \right) \,.
\ee
The exponent was computed in \cite{Hartnoll:2010gu} and in the small $\l$ limit becomes
\be
|\a| = \frac{3^{1/2} \hat m (2 - \hat m^2)^{1/2}}{(1 - \hat m^2)^{1/2}} \frac{1}{\l} - 1 - \frac{1}{2(1 - \hat m^2)} + \cdots \,.
\ee
The observation to make here is that the exponent $|\a|$ is becoming large as $\l \to 0$. This leads to an important simplification in the structure of the star. It means that the corrections to the IR Lifshitz behaviour are exponentially small for $r > 1$ (we will see shortly that the coefficients $f_1, g_1, h_1$ can be taken not to be also exponentially large or small in $\l$). Thus for $r > 1$ the Schr\"odinger equation (\ref{eq:schro}) takes the following Lifshitz form to leading order at small $\l$
\be\label{eq:s2}
\Phi'' =  \bar \g^2\left( \hat k^2 - \frac{\l^2 + 2 \hat m r^z \hat \w + r^{2z} \hat \w^2}{r^2}\right) \Phi \,,
\ee
where for convenience we introduced
\be
\bar \gamma^2 = \g^2 g_\infty^{(0)} \,.
\ee
In this last expression $g_\infty^{(0)}$ corresponds to the leading coefficient in the expansion of $g_\infty$ above.

In fact, more is true. The star ends also at $r=1$. This can be seen numerically or, alternatively, as follows. By solving the equations of motion \cite{Hartnoll:2010gu} perturbatively one finds that the coefficients appearing in the correction (\ref{eq:corrections}) can be taken to leading order in small $\lambda$ to be
\be\label{eq:fgh1}
f_1 = -1 \,, \qquad g_1 = - 2 \hat m^2 \,, \qquad h_1 = 
- \frac{2 \cdot 3^{1/2} \hat m^3 (1 - \hat m^2)^{1/2}}{(2 - \hat m^2)^{1/2}} \frac{1}{\l} \,.
\ee
The star ends at the radius $r = r_s$ where the potential in the Schr\"odinger equation (\ref{eq:schro}) vanishes at $\hat \w = \hat k = 0$. Using the first order corrections of (\ref{eq:fgh1}) it is easy to find that to leading order
\be
\frac{1}{r_s^{|\a|}} = \frac{1}{2 \hat m^2} \frac{1}{1-\hat m^2} \l^2 \,.
\ee
From this result we learn two things. Firstly that the correction in (\ref{eq:corrections}) is still small at the boundary of the star, and therefore our computation of $r_s$ is selfconsistent. Secondly, to leading order we have $r_s = 1$. Thus we find that the small $\lambda$ limit of electron stars has an interesting and simple structure: for $r > 1$ (inside the star) it is Lifshitz and for $r < 1$ (outside the star) it is Reissner-Nordstr\"om-AdS. This is potentially physically exciting as it means that all the fermions in the star are in the quantum critical geometry.

The local fermion density is therefore discontinuous at the edge of the star in the strict small $\l$ limit. At $r=1$ the density abruptly falls to zero. This kind of sharp endpoint of a `Fermi surface-like' object in the bulk is familiar from \cite{Rozali:2007rx, Shieh:2008nf} where instead of fermions there are endpoints of strings ending on a D brane and electrostatically repelling each other. A similar jump from Lifshitz to Reissner-Nordstr\"om is discussed in section 7.3 of \cite{Hartnoll:2009ns}. The reason that these discontinuities can arise is because we are looking for solutions subject to the constraint that there is a charge density at the boundary.

The fact that the boundary of the star occurs at $r_s=1$ is not a coordinate-invariant statement and ultimately depends on the choice of normalisation we made in (\ref{eq:fgh1}). We will later want to plot only physical, dimensionless ratios of quantities in which this ambiguity will cancel out. In order to do this, we need to compute the chemical potential of the dual field theory, $\hat \mu$, as well as the `speed of light', $c$. This can be obtained in the small star limit as follows.\footnote{This approach originated in collaboration with Pavel Petrov and was used to prove some assertions made in \cite{Hartnoll:2010ik}. Similar methods were also used in \cite{Puletti:2010de}.} Firstly note that the definitions of $\hat \mu$ and $c$ appearing in (\ref{eq:RNA}) can be extended away from the exterior Reissner-Nordstr\"om-AdS region by defining $c(r) = r^2 \sqrt{f g}$ and $c \hat \mu(r) = h - r h'$. The boundary values are then $c =c(0) = c(r_s)$ and $\hat \mu = \hat \mu(0) = \hat \mu(r_s)$. Differentiating these expressions, 
using the background equations of motion studied in \cite{Hartnoll:2010gu} and then integrating, one obtains
\be
c = \half \int_{r_s}^\infty r^3 h g^{3/2} \hat \sigma dr \,, \qquad c \hat \mu = \int_{r_s}^\infty \frac{r g \hat \sigma}{\sqrt{f}} \left(f - \half r h' h\right) dr \,.
\ee
In the small $\l$ limit all of these integrals may be evaluated on the Lifshitz background and with $r_s=1$. One obtains
\be\label{eq:cmu}
c = \frac{\hat \sigma h_\infty g_\infty^{3/2}}{2 (z-1)} \,, \qquad \hat \mu = \frac{g_\infty \hat \sigma}{c \, z} \left(1 + \frac{z h^2_\infty}{2}\right) \,.
\ee
Here the charge density $\hat \sigma$ is an order one constant in the small $\l$ limit. Using the definition of $\hat \sigma$ in \cite{Hartnoll:2010gu} one finds
\be
\hat \sigma = \frac{3^3 \hat m (1 - \hat m^2)}{2 \hat m^4 - 7 \hat m^2 + 6} + \cdots \,.
\ee

To start with it is instructive to consider the bound state spectrum. These are bound states of the Schr\"odinger problem (\ref{eq:s2}) with $\hat \w=0$. Solving this equation will determine the Fermi momenta $k_F^{(n)}$ of (\ref{eq:wkbn}) explicitly in the small star limit. At zero frequency it is easy to solve the Schr\"odinger equation subject to normalisability in the interior $r \to \infty$ to obtain the modified Bessel function
\be
\Phi = r^{1/2} K_{i \nu}(\bar \g \hat k r) \,.
\ee
Here the coefficient
\be\label{eq:nu}
\nu = \sqrt{\bar \g^2 \l^2 -\frac{1}{4}} \,.
\ee
The bound states are just given by imposing $\Phi=0$ at $r=1$, as the potential is exponentially large (at small $\lambda$) at $r<1$. The zeros of the Bessel function can be found explicitly in the small momentum limit $\bar \g \hat k \ll 1$ to be
\be\label{eq:kfn}
\bar \g \hat k^{(n)}_F = 2 e^{- (n + \frac{1}{2}) \frac{\pi}{\nu}} e^{\frac{1}{\nu} \text{Im} \log \Gamma(i \nu)} \,.
\ee
We see that the Fermi surfaces are accumulating exponentially at low momentum. This is because the Schr\"odinger potential has a maximum at the origin of the star. For these states to exist we need $\nu \in \R$ and thus, from (\ref{eq:nu}),
$2 \bar \g \l > 1 \,.$

\subsection*{Solvable case: z=2}

At finite frequencies the Schr\"odinger equation (\ref{eq:s2}) is still not solvable in general. However, it is clear that $z=2$ is a special value as only three different powers of $r$ (as opposed to four) appear in the Schr\"odinger equation (\ref{eq:s2}). When $z=2$ the solution to 
 (\ref{eq:s2}) satisfying ingoing boundary conditions at the horizon is the Tricomi confluent hypergeometric function
 \be\label{eq:hyper}
 \Phi = e^{i/2 \, r^2 \bar \g \hat \w} \, r^{1/2 + i \nu} \, U\left(\frac{1+i \nu}{2} + \frac{i (\hat k^2 - 2 \hat m \hat \w) \bar \g}{4 \hat \w}, 1+i \nu, - i r^2 \bar \g \hat \w \right) \,.
 \ee
Where $\nu$ is again as in (\ref{eq:nu}) and real. When $z=2$, to leading order at small $\l$, the mass $\hat m^2 = 1/2$, from e.g. (\ref{eq:zexp}). The quasinormal modes are found by imposing $\Phi(1)=0$. This condition can be solved explicitly in the low frequency limit $\bar \g \hat \w \ll 1$ to give
\be\label{eq:GG}
(- i \bar \g \hat \w)^{- i \nu} = - \frac{ \Gamma \left(\frac{1 + i \nu}{2} + \frac{i \bar \g (\hat k^2 - 2 \hat m \hat \w)}{4 \hat \w}  \right) \Gamma\left(- i \nu \right)}{ \Gamma \left(\frac{1 - i \nu}{2} + \frac{i \bar \g (\hat k^2 - 2 \hat m \hat \w)}{4 \hat \w}\right) \Gamma\left(i \nu \right)} \,.
\ee
The first computation we can perform with this result is to derive the dispersion of the fermion modes about the Fermi momenta (\ref{eq:kfn}). Taking the limit $\bar \g |(\hat k^2 - 2 \hat m \hat \w)/\hat \w| \gg 1$ of the above formula, one obtains the dispersions
\be\label{eq:z2pole}
\hat k^2 - \hat k_F^{(n)\,2} - 2 \hat m \hat \w = 0 \,.
\ee
This result is self consistent so long as $\bar \gamma^2 \hat k_F^{(n) \, 2} \gg \bar \gamma |\hat \w|$. We have therefore found the Fermi velocities of these poles.

The next step is to find the imaginary part of the dispersions. In fact simply by expanding (\ref{eq:GG}) it turns out that the pole (\ref{eq:z2pole}) never acquires an imaginary part at any order in the expansion in $\bar \g |(\hat k^2 - 2 \hat m \hat \w)/\hat \w| \gg 1$. This remains true if we keep higher order terms in the expansion of the confluent hypergeometric function in $\bar \gamma \hat \w$. To extract the exponentially small imaginary part we need to be more careful with the asymptotic series expansions of the gamma functions. In particular, we need to keep nonperturbative corrections to the usual expansion about Stirling's formula. Following \cite{berry} we can note that if we write
\be
\Gamma(z) = \sqrt{2 \pi} z^{z-\frac{1}{2}} e^{-z} e^{g(z)} \,,
\ee
then the reflection formula for gamma functions implies that
\be
g(z) = \frac{g(z) - g(-z)}{2} - \frac{1}{2} \log \left(1 - e^{2 \pi i z} \right) \,.
\ee
The first, antisymmetrized, term on the right hand side is the standard asymptotic expansion in terms of Bernoulli numbers. The second term gives corrections that are exponentially small in the upper half plane. Using the above formulae to refine the expansion of (\ref{eq:GG}) we find that the dispersion including the leading imaginary part is
\be
\hat k^2 - \hat k_F^{(n) \, 2} - 2 \hat m \hat \w + i \hat k_F^{(n) \, 2} \frac{\sinh \pi \nu}{\nu} e^{- \pi \bar \g \hat k_F^{(n) \, 2}/2 |\hat \w|} = 0 \,.
\ee
Thus for small electron stars we obtain explicitly the exponentially long lifetime that we discussed more generally around (\ref{eq:twofinal}) above.

As well as exponentially stable poles, the expression (\ref{eq:GG}) captures the poles that are away from the real axis, some of which we discussed in section \ref{sec:wktext} above. A virtue of the exact solution (\ref{eq:hyper}) is that it interpolates between cases II and III without needing to perform separate WKB matching procedures. In particular, it captures all the poles in the Green's function in the single expression (\ref{eq:GG}). This expression is similar to the poles of formula (\ref{eq:II3}) for the boundary between cases II and III, except that (\ref{eq:GG}) has a wider regime of validity. To make these statements precise, we proceed to find the boundary Green's function.

To find the Green's function, we need to match the solution of the Dirac equation near the boundary of the star onto a solution in the exterior region. To connect with the rest of this paper, we should do this in the strict WKB limit. As we discuss in appendix \ref{sec:ee} this involves taking $\nu \sim \bar \g \lambda \gg 1$. Expanding (\ref{eq:hyper}) as previously in $r \bar \g \hat \w \ll 1$, using (\ref{eq:c1}) and (\ref{eq:c2}) to match the resulting solution onto an exterior WKB solution and then using (\ref{eq:aa2}) to read off the Green's function, we obtain:
\be\label{eq:green}
G^R(\hat \w,\hat k) = \frac{i \hat \mu}{4 \hat m} e^{- 2 \g I_B} \Bigg(1 -
\frac{2 (-i \bar \g \hat \w)^{- i \nu}}{ (-i \bar \g \hat \w)^{- i \nu} - e^{i \pi/2}
 \frac{ \Gamma \left(\frac{1 + i \nu}{2} + \frac{i \bar \g (\hat k^2 - 2 \hat m \hat \w)}{4 \hat \w}  \right) \Gamma\left(- i \nu \right)}{ \Gamma \left(\frac{1 - i \nu}{2} + \frac{i \bar \g (\hat k^2 - 2 \hat m \hat \w)}{4 \hat \w}\right) \Gamma\left(i \nu \right)}} \Bigg) \,.
\ee
We see that the poles of the Green's function essentially agree with (\ref{eq:GG}).
The extra factor of $e^{i \pi/2}$ is the usual shift in the zero point energy that, while subleading, is captured correctly in a WKB treatment. The boundary condition we used above, setting the wavefunction to zero at $r=1$, was too crude to capture this correction.\footnote{A linear WKB matching across this turning point means we cannot access near extermal Fermi momenta between cases I and II. However we still capture the crossover from cases II to III, our primary interest.} Perhaps curiously, the Green's function (\ref{eq:green}) bears some resemblance to the IR Green's functions of extremal Reissner-Nordstr\"om-AdS \cite{Faulkner:2009wj}. In the prefactor of (\ref{eq:green}) we used the fact that $\hat \w, \hat k \ll \hat \mu$ in the regime we will be interested in. The exponent in the overall suppression $e^{- 2 \g I_B}$ due to tunneling from the boundary is again given by (\ref{eq:iib}) with $r_1=1$. The frequency and momentum dependence of this term will be weak compared to the remainder of the Green's function and so this term is taken to be a constant in the plots below.
\begin{figure}[h]
\begin{center}
\includegraphics[width=213pt]{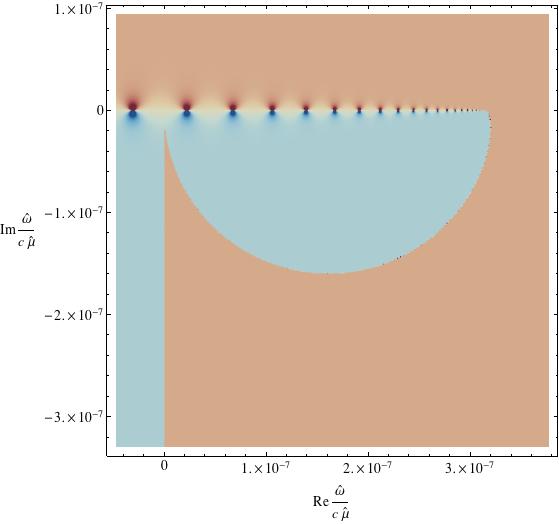}\hspace{0.3cm}\includegraphics[width=210pt]{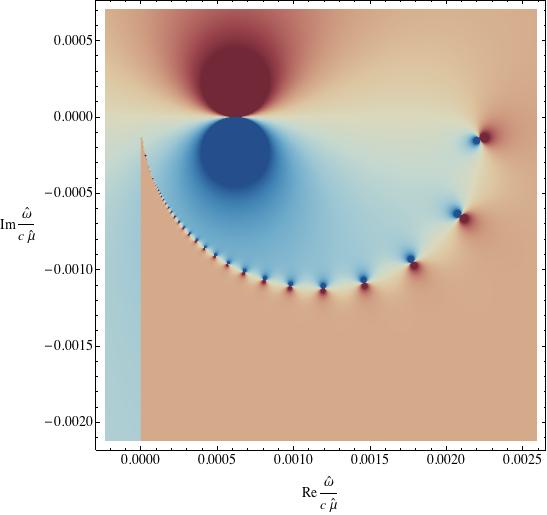}\caption{Poles in the retarded Green's function: Imaginary part of the retarded Green's function, $\text{Im} G^R (\hat \w,\hat k)$, as a function of complex frequency at fixed real momentum. Computed using (\ref{eq:green}) with $\hat m=1/\sqrt{2}$, corresponding to $z=2$ as required. Left plot has $\bar \gamma = 1000, \nu = 40$ and $\hat k/\hat \mu = 2/3000$. Right plot has $\bar \gamma = 10, \nu = 4$ and $\hat k/\hat \mu = 2/30$.  \label{fig:prettypictures}}
\end{center}
\end{figure}

In plotting the Green's function we should use dimensionless variables. For $z=2$ we find from the formulae in the previous subsection, such as (\ref{eq:cmu}), that $c=\sqrt{2}$ and $\hat \mu = 3/2$. Plotting the Green's function against $\hat \w/(c \hat \mu)$ and $\hat k/\hat \mu$ will undo the coordinate choice we made in taking $r_s=1$ while also fixing the normalisation of time of the boundary quantum field theory so that the speed of light is unity. The result is shown in figure \ref{fig:prettypictures}.

In figure \ref{fig:prettypictures} we see several features discussed previously. There are poles along the real frequency axis corresponding to excitations of the many different Fermi surfaces, but beyond $\hat \w \sim \hat k^2$ the spectral function is constant. The poles away from the real axis form the semicircle anticipated in section \ref{sec:wktext}. To understand the semicircle we can relate the poles of the small star Green's function (\ref{eq:green}) to the general Green's function (\ref{eq:II3}) close to the dispersion relation $\hat \w \sim \hat k^2$. To connect the two we must take the large $\nu$ limit of (\ref{eq:green}) keeping $\ep$ fixed. Of the two more complicated gamma functions appearing in the denominator of (\ref{eq:green}), the one in the numerator is $\G(\frac{1}{2} - \ep + i \nu)$ while the one in the denominator is $\G(\frac{1}{2} - \ep)$. At large $\nu$ with $\ep$ fixed we land on the general result (\ref{eq:II3}). Repeating the arguments of section \ref{sec:wktext} gives a semicircle of poles in the lower half plane.

The semicircles form essentially a right angle with the real axis in figure \ref{fig:prettypictures}, different from the small angle of section \ref{sec:wktext}. In the small star limit the angle is determined by $\log \g \lambda \sim \log \nu$ rather than the $\log \g$ of equation (\ref{eq:ppp}), see appendix \ref{sec:ee}. For the values of the parameters chosen in the plots, this logarithm is not sufficiently large to lead to a small angle.

Because the Green's function (\ref{eq:green}) captures both cases II and III, we can plot $\text{Im} G^R(\hat \w, \hat k)$ to produce a precise version of the sketch of figure \ref{fig:wkplane}. The result is shown in figure \ref{fig:spectraldensity}.
\begin{figure}[h]
\begin{center}
\includegraphics[width=190pt]{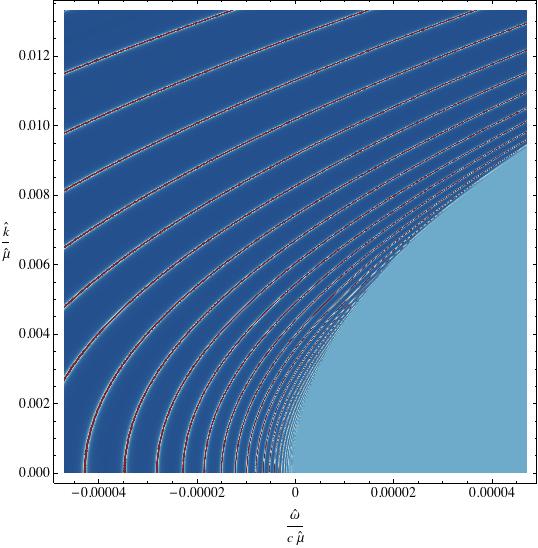}\caption{Spectral density of many Fermi surfaces interacting with a dissipative critical sector: $\text{Im} G^R (\hat \w,\hat k)$, as a function of frequency and momentum. Computed using (\ref{eq:green}) with $\hat m=1/\sqrt{2}$, corresponding to $z=2$ as required. The plot has $\bar \gamma = 100, \nu = 30$.  \label{fig:spectraldensity}}
\end{center}
\end{figure}

As a final plot, we illustrate more explicitly how the spectral weight of the stable Fermi surface poles are dissipated away as they pass through the critical dispersion relation (\ref{eq:disper}). Figure \ref{fig:dis} shows the spectral density of a pole along the real frequency axis as the momentum of the pole moves into the unstable region.
\begin{figure}[h]
\begin{center}
\includegraphics[width=220pt]{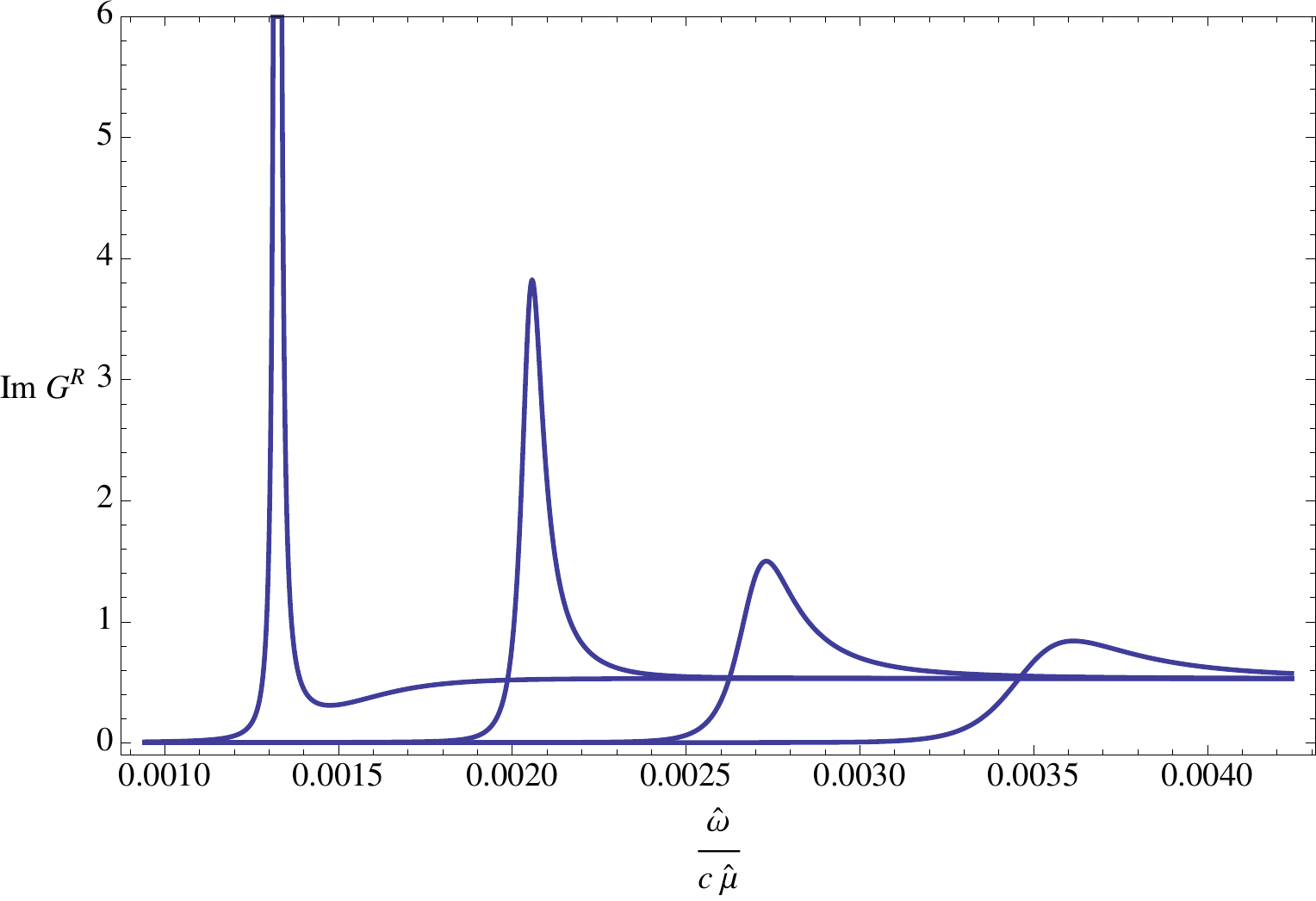}\caption{Dissolution of the Fermi surface pole in the critical region: Spectral density as a function of frequency for various values of momentum, from $\hat k/\hat \mu = 0.053$ for the leftmost peak to $\hat k/\hat \mu = 0.08$ for the rightmost. The plot has $\bar \gamma = 10, \nu = 3$.  \label{fig:dis}}
\end{center}
\end{figure}

\section{Discussion}

In this paper we have characterised the Green's function $G^R(\w,k)$ of gauge-invariant fermion operators in phases of 2+1 dimensional field theories at finite charge density dual to electron stars. Three main results are
\begin{itemize}

\item $G^R(0,k)$ shows a large number of closely spaced Fermi surface for $k < k_F^\star$. The volumes of these Fermi surfaces add up to the charge density, thereby satisfying the 2+1 dimensional Luttinger count.

\item $G^R(\w,k)$ changes qualitatively at a specific dispersion: $\w \sim k^z$. For real $\w \lesssim k^z$ the Fermi surface excitations are very stable, while for $\w \gtrsim k^z$ they are unstable and the spectral density eventually becomes featureless.

\item Smaller $k_F$ Fermi surfaces have exponentially larger spectral weight than the more stable large $k_F$ Fermi surfaces.

\end{itemize}
We also estimated that the critical fermions near the dispersion $\w \sim k^z$ may dominate some physical observables such as the optical conductivity.

As emphasised in \cite{deBoer:2009wk, Arsiwalla:2010bt, Hartnoll:2010xj}, electron (and neutron) stars in the bulk should be thought of as Fermi surfaces of generalised free fields. Generalised free fields (see e.g. \cite{ElShowk:2011ag}) are the mechanism whereby holography is able to give a semiclassical description of strongly interacting physics: the spectral weight of an operator in the field theory is distributed among an infinite tower of generalised free fields, which correspond to different harmonics of the radial quantm number in the bulk. Thus, the large number of closely spaced Fermi surfaces can be thought of as a `smeared' Fermi surface of the strongly interacting fermion operator \cite{Hartnoll:2010xj}. We have seen explicitly in this paper that WKB resolves the individual Fermi surfaces. Whether or not the spectral weight appears smeared or discrete will depend on the observable.

The fact that the field theory Luttinger count works out for electron stars indicates that they are not dual to `fractionalised' Fermi liquids in the sense of \cite{fraction, Huijse:2011hp}. Nonetheless, there is no indication that the field theory $SU(N)$ gauge symmetry is Higgsed in these backgrounds, implying that the transverse gauge field excitations should remain gapless at the lowest energy scales. The IR Lifshitz geometry of the electron star should then be thought of as a gauge-invariant and strong coupling cousin of Landau damping of the gauge field excitations by the density of fermions. Many of the excitations of the Fermi surfaces are unable to dissipate efficiently into the critical Lifshitz modes. The appearance of a particular dispersion $\w \sim k^z$, separating regimes that can and cannot dissipate, has the flavor of a `non-relativistic lightcone', suggesting an emergent kinematic constraint controlling the IR interactions between bosons and fermions. We feel that this dispersion relation remains to be further understood.

It is of interest to find gravity duals to fractionalised Fermi liquid phases. These will be characterised by the fact that the total volume of bulk gauge-invariant Fermi surfaces will add up to less than the total charge density (and that there is at least one such Fermi surface). This indeed occurs if one neglects the backreaction of the bulk fermions on the geometry and studies fermion correlators in the extremal Reissner-Nordstr\"om-AdS background as in \cite{Liu:2009dm, Cubrovic:2009ye, Faulkner:2009wj}. However, even for the fermions considered in those papers, where the effects of backreaction are subleading in the bulk classical limit, it is plausible that once backreaction of the bulk Fermi surface is incorporated then all the charge is carried by fermions rather than the black hole horizon \cite{Hartnoll:2009ns}. This will then likely lead to the Luttinger count being satisfied, as we found in this paper for the electron star. The conclusion, in that case, would be that the `large $N$' and low energy limits do not commute in terms of diagnosing the phase of matter. The electron star limit was designed to have the virtue of capturing the correct low energy physics at leading order in the large $N$ (i.e. bulk classical) expansion \cite{Hartnoll:2010gu}. The objective is therefore to find a context where a backreacting bulk Fermi surface can coexist with a charge-carrying horizon at zero temperature. The recently discussed Fermi surfaces in dilaton gravity \cite{Iizuka:2011hg} could be one interesting case to consider.

\section*{Acknowledgements}

It is a pleasure to thank Alireza Tavanfar for collaboration in the initial stages of this project. We have also benefitted from discussions with Liza Huijse, John McGreevy, Max Metlitski and Subir Sachdev while this work was in progress. The research of S.A.H. is partially supported by DOE grant DE-FG02-91ER40654 and the FQXi foundation. D.M.H. would like to thank the Center for the Fundamental Laws
of Nature at Harvard University for support.

\appendix

\section{Intermediate matching formulae for cases I, II and III}
\label{sec:matchingeasy}

To start, recall the well known matching conditions across a generic turning point where $V$ vanishes linearly in radius. Then, the Schr\"odinger equation is solved in the matching region by Airy functions. Schematically, with $x$ changing sign across the turning point, we have
\bea
\frac{1}{2} e^{- x^{3/2}} \quad \leftrightarrow \qquad  \sin \left((-x)^{3/2} + \pi/4 \right) \,, \label{eq:c1} \\
e^{x^{3/2}} \quad \leftrightarrow \qquad \cos \left((-x)^{3/2} + \pi/4 \right) \,. \label{eq:c2}
\eea
As we noted in the discussion below equation (\ref{eq:schro}), from the start we are neglecting terms that have effects comparable to the $\pi/4$ phase shifts here. Nonetheless we will keep track of the phase shifts with a view to solving the WKB problem posed by equation (\ref{eq:schro}) precisely.

At the horizon we must impose ingoing boundary conditions. Therefore, in the region between $r=\infty$ and the first turning point ($r_3$ in cases I and II, $r_1$ in case III) we must have when $\w > 0$ that
\be
\Phi_{r > r_{1/3}} = \exp\left\{i \, \g \int^r_{r_{1/3}} dr \sqrt{-V}  \right\} \,.
\ee
Using the connection formulae (\ref{eq:c1}) and (\ref{eq:c2}) we find that on the other side of the matching point
\be
\Phi_{r < r_{1/3}} = e^{-i \pi/4} \left( \exp\left\{\g \int^{r_{1/3}}_r dr \sqrt{V}\right\} + \frac{i}{2} \exp\left\{- \g \int^{r_{1/3}}_r dr \sqrt{V} \right\}  \right) \,. 
\ee
In the cases I and III this region beyond the first matching point is already the outer region. Comparing with expressions (\ref{eq:pm}), (\ref{eq:pm1}) and (\ref{eq:pm2}) we can conclude that the retarded Green's function at leading order in the WKB expansion is
\be\label{eq:13}
G^R(\w,k) \propto \frac{i}{2} \lim_{r \to 0} r^{- 2 m L} \exp\left\{- 2 \g \int^{r_{1/3}}_r dr \sqrt{V} \right\}  \,, \qquad \text{(cases III and I)} \,.
\ee
This was formula (\ref{eq:13text}) in the main text.

In case II we need to propagate the solution through the intermediate regions. Using the connection formulae (\ref{eq:c1}) and (\ref{eq:c2}) two additional times, it is straightforward to obtain
\be\label{eq:two}
G^R(\w,k) \propto \frac{i \, {\mathcal G}}{2} \lim_{r \to 0} r^{- 2 m L} \exp\left\{- 2 \g \int^{r_1}_r dr \sqrt{V} \right\}  \,, \qquad \text{(case II)}
\ee
where we specify ${\mathcal G}$ in the formulae below (\ref{eq:twotext}) in the main text.

\section{Matching formulae for $k \sim k_F^\star$}
\label{sec:kf}

This appendix contains the computation of the matching across a minimum of the potential necessary for section \ref{sec:kftext}. Firstly, we should derive equation (\ref{eq:AB2}) which, combined with equation (\ref{eq:AB}), describes the potential (\ref{eq:potential}) near the minimum. To zoom in on the problematic region near $r = r_\star$, we let $\hat \w = \d \hat \w$, $\hat k = \hat k_F^\star + \d \hat k$ and $r = r_\star + \d r$. Expanding the potential (\ref{eq:potential}) in all three variables gives (to first order in the frequency and momentum and to second order in $\d r$)
\bea\label{v2}
V(r) & = & 2 \left(\hat k_F^\star \tilde g(r_\star) \d \hat k - j(r_\star) \d \hat \w\right) + 2 \left(\hat k_F^\star \tilde g'(r_\star) \d \hat k - j'(r_\star) \d \hat \w\right) \d r \nonumber \\
& & \, + \left(\hat k_F^\star \tilde g''(r_\star)\d \hat k - j''(r_\star) \d \hat \w - \hat k_F^\star k_F''(r_\star) \tilde g(r_\star)\right) \d r^2 + \cdots \,. 
\eea
In order to keep the length of the formula down, we defined $\tilde g(r) = r^2 g(r)$ and $j(r) = g(r) h(r)/f(r)$.
From this last expression we see that turning on the variations $\delta \hat \omega, \delta \hat k$ moves the extremum from $\delta r = 0$. The new extremum is at
\be
\delta r_\star = \frac{\hat k_F^\star  \tilde g'(r_\star)  \delta \hat k -  j'(r_\star) \delta \hat \omega  }{\tilde g(r_\star) \hat k_F^\star \hat k_F''(r_\star)} \,,
\ee
to leading order in $\delta \hat \omega, \delta \hat k$. We wish to follow the behavior near the extremum, so we will consider this an expansion around the actual minimum $r_o = r_\star + \delta r_\star$. Substituting into (\ref{v2}) and re-expanding about the true minimum the potential becomes
\be
V(r) = 2 \left( \hat k_F^\star  \tilde g(r_\star) \delta \hat k - j(r_\star)  \delta \hat \omega \right)  -  \tilde g(r_\star) \hat k_F^\star \hat k_F''(r_\star) (r - r_o)^2 + \cdots \,.
\ee
Recall that since $\hat k_F^\star$ is a maximum of $\hat k_F(r)$,  $\hat k_F''(r_\star)$ is negative. Comparing with the quadratic potential (\ref{eq:AB}) in the main text we can read off the values of $A$ and $B$ quoted in (\ref{eq:AB2}).

In the main text we then performed the rescaling (\ref{eq:rescale}) in order to put the Schr\"odinger equation in the form (\ref{eq:Vmin}) which we rewrite here for ease of reference:
\be\label{eq:VminA}
- \ddot \Phi + \left(-\e + \frac{y^2}{4} \right) \Phi = 0 \,. 
\ee
We can now solve this potential exactly and describe how we use the exact solution to perform matching. The general solution is
\be
\Phi(y) = C_1 D_{-\frac{1}{2} + \e}(y) + C_2 D_{-\frac{1}{2}-\e}(i y) \,,
\ee
where $D_s(y)$ are the parabolic cylinder $D$ functions of order $s$. The asymptotic expansion of these functions can now be used to match onto the WKB solutions. The common region where both the WKB and the quadratic potential approximations are valid is when $\max(1, \sqrt{\e}) \ll |y| \ll \sqrt{\g}$.
Expanding for large $|y|$ we can quote the asymptotic expansions of $\Phi$:
\begin{eqnarray}\label{efunc}
\Phi(y)_{y \rightarrow + \infty}  & \longrightarrow & C_1  e^{-\frac{y^2}{4}} y^{-\frac{1}{2}+\e} + \left( e^{-i \e \frac{\pi}{2}} e^{-i\frac{\pi}{4}} C_2\right) e^{+\frac{y^2}{4}} y^{-\frac{1}{2}-\e} \,, \\
\Phi(-y)_{y \rightarrow + \infty}  & \longrightarrow & \left( e^{i \e \pi} e^{- i \frac{\pi}{2}} C_1 +\frac{e^{i \e \frac{\pi}{2}}e^{-i \frac{\pi}{4}}\sqrt{2\pi}}{\Gamma\left(\frac{1}{2}+\e\right)} C_2 \right) e^{-\frac{y^2}{4}} y^{-\frac{1}{2}+\e} \nonumber\\
& &+ \left(\frac{\sqrt{2\pi}}{\Gamma\left(\frac{1}{2}-\e\right)} C_1+ e^{i \e \frac{\pi}{2}} e^{i \frac{\pi}{4}} C_2\right) e^{+\frac{y^2}{4}} y^{-\frac{1}{2}-\e} \,. \label{efunc2}
\end{eqnarray}

Now we should match these results with the expansions of the WKB wave functions. The WKB wavefunction for $\delta r \equiv r - r_o >0$, that is, towards the star interior, takes the form
\be\label{eq:phir}
\Phi(r) = \frac{R_+}{(\g^2 V)^{\frac{1}{4}}} e^{+\g \int^{r_3}_{r_o+\delta r} \sqrt{V(r)} \,dr} +\frac{R_-}{(\g^2 V)^{\frac{1}{4}}} e^{-\g \int^{r_3}_{r_o+\delta r} \sqrt{V(r)} \,dr} \,.
\ee
In order to evaluate the integrals in the exponents here, it is useful to write, recalling the form of the potential (\ref{eq:AB}), $V(r) = -A + \Delta V(r)$. We can then note that given that $r_o + \delta r$ is in the overlap region ($\sqrt{A} \ll \d r \ll 1$), then we have $A \ll \Delta V(r)$ over the whole range of integration. Expanding to first order in $A$ we can then write
\be\label{eq:adap}
\int^{r_3}_{r_o+\delta r} \sqrt{V(r)} \,dr = \int_{r_o}^{r_3} \left( \sqrt{\Delta V} - \frac{A}{2 \sqrt{\Delta V}} + \frac{A}{B |r-r_o|} \right) dr - \frac{B (\d r)^2}{4} -
\frac{A}{B} \log \frac{r_3 - r_o}{\d r} \,.
\ee
Note that a potential logarithmic divergence cancels between the second and third terms in the integral.

Using the rescaling (\ref{eq:rescale}) we can then express the WKB solution (\ref{eq:phir}) for $\d r > 0$ in the matching region as
\be\label{eq:phir2}
\Phi = \frac{\sqrt{2} R_+}{(\g B)^{\frac{1}{4}}} e^{\gamma I_S + \e T_S} y^{-\frac{1}{2}+ \e} e^{-\frac{y^2}{4}} +  \frac{\sqrt{2} R_-}{(\g B)^{\frac{1}{4}}} e^{-\gamma I_S - \e T_S} y^{-\frac{1}{2}- \e} e^{+\frac{y^2}{4}} \,.
\ee
where 
\begin{eqnarray}
I_S &=&  \int^{r_3}_{r_o} \sqrt{\Delta V} \,dr  \,, \\
T_S &=& \int^{r_3}_{r_o}\left( - \frac{B}{2 \sqrt{\Delta V}} + \frac{1}{ |r-r_o|} \right) \,dr -  \log{ \sqrt{\gamma B}  |r_3-r_o|} \,.
\end{eqnarray}
The subscript $S$ is supposed to remind us these are integrals calculated over the repulsion generated by the star. Notice that these quantities are not a function of our Schr\"odinger energy $\e$.

Equating (\ref{efunc}) and (\ref{eq:phir2})
\begin{eqnarray}\label{cr}
C_1 &=& \frac{\sqrt{2} R_+}{(\g B)^{\frac{1}{4}}} e^{\gamma I_S + \e T_S} \,, \\
C_2 &=&  \frac{\sqrt{2} R_-}{(\g B)^{\frac{1}{4}}}  e^{i \e \frac{\pi}{2}} e^{i\frac{\pi}{4}} e^{-\gamma I_S - \e T_S} \,.
\end{eqnarray}
This is one half of the matching procedure. We can now go through the same exercise on the other side of the minimum of the potential. The solutions in the matching regime for $\delta r \equiv r_o - r >0$ are:
\begin{eqnarray}\label{rfunc}
\Phi &=& \frac{L_+}{(\g^2 V)^{\frac{1}{4}}} e^{+\g \int_{0}^{r_o-\delta r} \sqrt{V(r)} \,dr} +\frac{L_-}{(\g^2 V)^{\frac{1}{4}}} e^{-\g \int_{0}^{r_o-\delta r} \sqrt{V(r)} \,dr} \\
&=& \frac{\sqrt{2} L_+}{(\g B)^{\frac{1}{4}}} e^{\gamma I_B + \e T_B} |y|^{-\frac{1}{2}+ \e} e^{-\frac{y^2}{4}} +  \frac{\sqrt{2} L_-}{(\g B)^{\frac{1}{4}}} e^{-\gamma I_B - \e T_B} |y|^{-\frac{1}{2}- \e} e^{+\frac{y^2}{4}} \,, \label{eq:pp}
\end{eqnarray}
where now
\begin{eqnarray}
I_B &=&  \int_{0}^{r_o} \sqrt{\Delta V} \,dr \,, \\
T_B &=& \int_{0}^{r_o}\left( - \frac{B}{2 \sqrt{\Delta V}} + \frac{1}{ |r-r_o|} \right)\,dr -  \log{ \sqrt{\gamma B}  |r_o|} \,.
\end{eqnarray}
Here the subscript $B$ is reminding us that these are integrals to the boundary of $AdS$. One small comment is that integrals diverge near the boundary; rather than integrating to $r=0$ we should introduce a UV cutoff as in (\ref{eq:pm}). Ultimately, in the Green's function, the integral will be regularised as in (\ref{eq:twotext}). We incorporated the regularisation terms in the corresponding equations in the main text.

Matching the result (\ref{eq:pp}) with (\ref{efunc2}) we obtain
\begin{eqnarray}
L_+ &=& \frac{\left(\gamma B\right)^{\frac{1}{4}}}{\sqrt{2}} e^{-\gamma I_B - \e T_B}  \left(e^{i \e \pi} e^{- i \frac{\pi}{2}} C_1 +\frac{e^{i \e \frac{\pi}{2}}e^{-i \frac{\pi}{4}}\sqrt{2\pi}}{\Gamma\left(\frac{1}{2}+\e\right)} C_2 \right) \,, \\
L_- &=& \frac{\left(\gamma B\right)^{\frac{1}{4}}}{\sqrt{2}} e^{+\gamma I_B + \e T_B} \left(\frac{\sqrt{2\pi}}{\Gamma\left(\frac{1}{2}-\e\right)} C_1+ e^{i \e \frac{\pi}{2}} e^{i \frac{\pi}{4}} C_2\right) \,.
\end{eqnarray}
Finally, we can now eliminate the matching region altogether by using the above expression together with (\ref{cr}). The coefficients of the WKB solution on either side of the matching region are thereby related as follows:
\begin{eqnarray}
L_+ &=& e^{-\gamma I_B - \e T_B} e^{i \e \pi} \left(e^{- i \frac{\pi}{2}}  R_+ e^{\gamma I_S + \e T_S}  +\frac{\sqrt{2\pi}}{\Gamma\left(\frac{1}{2}+\e\right)}  R_- e^{-\gamma I_S - \e T_S}  \right) \,, \\
L_- &=& e^{+\gamma I_B + \e T_B} \left(\frac{\sqrt{2\pi}}{\Gamma\left(\frac{1}{2}-\e\right)}  R_+ e^{\gamma I_S + \e T_S} + e^{i \e \pi} e^{i \frac{\pi}{2}}  R_- e^{-\gamma I_S - \e T_S} \right) \,.
\end{eqnarray}

Recall, using the linear matching conditions (\ref{eq:c1}) and (\ref{eq:c2}), that imposoing ingoing boundary conditions at the horizon and matching across the turning point $r_3$ nearest the horizon will set $R_- = \frac{e^{i \frac{\pi}{2}}}{2} R_+$. Thus we have all the ingredients to find the Green's function. Taking $r \to 0$ in (\ref{rfunc}) we have
\be
G^R(\w,k) \propto \lim_{r \to 0} r^{2 m L} \frac{L_+}{L_-} \,.
\ee
The regularising $r^{2 m L}$ can be incorporated into the definition of $I_B$ and $T_B$ and we have done this in the main text.
From our above results
\be\label{green1}
\frac{L_+}{L_-} = e^{-2\gamma I_B - 2\e T_B} e^{i \e \pi}  e^{- i \frac{\pi}{2}} \frac{1 -  \frac{1}{2} \frac{\sqrt{2 \pi}}{\Gamma\left(\frac{1}{2}+\e \right)}e^{-2\gamma I_S - 2\e T_S}}{\frac{\sqrt{2\pi}}{\Gamma\left(\frac{1}{2}-\e \right)} - \frac{1}{2}e^{i \e \pi}  e^{-2\gamma I_S - 2\e T_S} } \,.
\ee
This is the Green's function we have quoted in equation (\ref{eq:green1text}) in the main text.

\section{Matching formulae for $k \sim \w^{1/z}$}
\label{sec:wk}

The Schr\"odinger equation in the regime across which we wish to match is
\be
- \ddot \Phi + \left(-\e - \frac{y^2}{4} \right) \Phi = 0 \,. 
\ee
This is exactly the same problem as we have solved in the previous appendix, equation (\ref{eq:VminA}), but with 
 $y \rightarrow \sqrt{i} y$ and $\e \rightarrow - i \e$. Therefore, the solutions are
\be
\Psi(y) = C_1 D_{-\frac{1}{2} - i \e}(e^{i \frac{\pi}{4}} y) + C_2 D_{-\frac{1}{2}+ i \e}(e^{i \frac{3 \pi}{4}}  y) \,.
\ee
It is important to be careful about the asymptotic expansions here, as the rotation in the complex plane does not allows us to just analytically continue the expansions of the $D$ functions we used before. In the present case
\begin{eqnarray}\label{eIfunc}
\Psi(y)_{y \rightarrow + \infty}  &\longrightarrow& \left(\frac{e^{i \frac{\pi}{8}} e^{-\e\frac{\pi}{4}} \sqrt{2 \pi}}{\Gamma\left(\frac{1}{2} - i \e\right)} C_2 -   e^{ \e \frac{\pi}{4}} e^{i\frac{7 \pi}{8}} C_1\right)  e^{-i \frac{y^2}{4}} y^{-\frac{1}{2}- i\e}\nonumber\\
&+& \left( -  e^{- \e \frac{3\pi}{4}} e^{i\frac{5\pi}{8}} C_2 \right) e^{+i \frac{y^2}{4}} y^{-\frac{1}{2}+ i\e} \,, \\
\Psi(-y)_{y \rightarrow + \infty}  &\longrightarrow& \left( e^{- \e \frac{3 \pi}{4}} e^{i \frac{3\pi}{8}} C_1 \right) e^{-i \frac{y^2}{4}} y^{-\frac{1}{2}-i \e} \nonumber\\
& &+ \left(  e^{\e \frac{\pi}{4}} e^{i \frac{\pi}{8}} C_2 -  \frac{\sqrt{2\pi} e^{-\e \frac{\pi}{4}} e^{i\frac{7 \pi}{8}} }{\Gamma\left(\frac{1}{2}+ i \e\right)}C_1 \right)e^{+i \frac{y^2}{4}} y^{-\frac{1}{2}+ i \e} \,. \label{eIfunc2}
\end{eqnarray}

For positive $y$ we need the wave function to match onto a mode that is falling into the horizon. This means the coefficient of the $e^{-i \frac{y^2}{4}}$ term needs to vanish. This requires
\be
C_2\frac{e^{i \frac{\pi}{8}} e^{-\e\frac{\pi}{4}} \sqrt{2 \pi}}{\Gamma\left(\frac{1}{2} - i \e\right)} - C_1  e^{ \e \frac{\pi}{4}} e^{i\frac{7 \pi}{8}}= 0 \,.
\ee
From this we obtain (up to an inessential normalization) the form of the wave function at negative $y$ to be
\be\label{eq:ybigcase2}
\Psi(-y)_{y \rightarrow + \infty}  \longrightarrow \left( i e^{\e \frac{\pi}{2}} \frac{\sqrt{2 \pi}}{\Gamma\left(\frac{1}{2} - i \e\right)} \right) e^{-i \frac{y^2}{4}} y^{-\frac{1}{2}-i \e} + e^{i \frac{y^2}{4}} y^{-\frac{1}{2}+i \e} \,.
\ee
If we were looking for eigenstates with purely outgoing boundary condition we would obtain $\e = - i (\frac{1}{2}+n)$, the analytic continuation of the standard harmonic oscillator problem. Here, however, we need to match this wavefunction into a WKB approximation that captures the fact that particles can oscillate inside the potential well.
The WKB solution takes the form
\be
\Phi = \frac{Q_+}{(- \g^2 V)^{\frac{1}{4}}} e^{+ i \g \int_{r_1}^{r_o-\d r} \sqrt{-V} dr} + \frac{Q_-}{(- \g^2 V)^{\frac{1}{4}}} e^{- i \g \int_{r_1}^{r_o-\d r} \sqrt{-V} dr} \,.
\ee
To match this WKB result onto the expansion (\ref{eq:ybigcase2}) we use the analogous representation to that in (\ref{eq:adap}), letting $V = - A + \Delta V$. To first order in $A$
\be\label{regintI}
\int_{r_1}^{r_o- \d r} \sqrt{-V} dr = \int_{r_1}^{r_o} \left(\sqrt{-\D V} + \frac{A}{2\sqrt{-\D V}} - \frac{A}{B |r - r_0|} \right)dr - \frac{B (\d r)^2}{4} - \frac{A}{B} \log \frac{\d r}{r_o - r_1} \,.
\ee
Recalling the rescalings (\ref{eq:rescale}), matching leads to
\be
\frac{Q_+}{Q_-} = \frac{i e^{\ep \frac{\pi}{2}} \sqrt{2\pi}}{\G\left(\frac{1}{2} - i \ep\right)} e^{- 2 i \g I_W - 2 i \ep T_W} \,,
\ee
where
\begin{eqnarray}
I_W &=& \int_{r_1}^{r_o} \sqrt{- \D V} \, dr \,, \label{eq:iw} \\
T_W &=& \int_{r_1}^{r_o} \left( \frac{B}{2 \sqrt{-\D V} } - \frac{1}{|r-r_o|} \right) \, dr + \log{\sqrt{\g B} |r_o-r_1|} \,. \label{eq:tw}
\end{eqnarray}
where $W$ is supposed to remind us that we are integrating over the well.

The final step in computing the Green's function is to match the oscillating WKB region onto the asymptotic regime. This is a standard linear WKB matching, using (\ref{eq:c1}) and (\ref{eq:c2}). From the definition (\ref{eq:aa2}) we obtain
\be\label{eq:grappend}
G^R = \frac{\hat \mu}{2 \hat m} \frac{i}{2} \frac{\frac{Q_+}{Q_-} - i}{\frac{Q_+}{Q_-} + i} e^{-2 \gamma I_B} \,,
\ee
where $I_B$ is again the regularized integral characterizing the amplitude to tunnel in from the boundary. It is given by
\be
I_B =  \int_{0}^{r_1} \left( \sqrt{V} - \frac{\hat m}{r} \right) \,dr + \hat m \log r_1 \,. 
\ee
The above Green's function is the result we quote in the main text in (\ref{eq:II3}).

It is instructive to identify the poles in the Green's function (\ref{eq:grappend}). These occur at
\be
- e^{\e \frac{\pi}{2}} \frac{\sqrt{2 \pi}}{\Gamma\left[\frac{1}{2} - i \e\right]}  = e^{i 2\gamma I_W +  i 2 \e T_W} \,.
\ee
There are two types of solutions to this equation. Most of the poles can be captured using Stirling's approximation for
the gamma function, i.e. taking $|\e| \gg 1$. Explicitly separating into real and imaginary parts:
\be
\e = \e_1 + i \e_2 \,,
\ee
and taking the logarithm of equation for the poles, we obtain
\begin{eqnarray}
\e_1 \left(\pi-\arg(\e)\right)- \e_2 \left( \log{|\e|}-1\right) &+& i \left[ \e_1\left( \log{|\e|}-1\right)  + \e_2 \left(\pi-\arg(\e)\right)\right]\nonumber\\
&=& i \pi (2 n+1) + i 2\gamma I_W +  i 2 \e T_W \,.\label{eq:polesexp}
\end{eqnarray}
Consider the real and imaginary parts of this equation. Firstly we can see that there are solutions with $\e_2 = 0$ and $\e_1 < 0$ if $I_W$ and $T_W$ are real. These poles are therefore along the negative real $\e_1$ axis and have
\be\label{reale}
\e_1 \left(\log{|\e_1|} -1 -2 T_W\right) =\pi (2n+1) + 2 \gamma I_W \,,
\ee
where $n$ is constrained to run over all integers such that $\e_1<0$. Working a little harder with the expansion of the gamma function, and we do so in section \ref{sec:smallstar}, one can show that these poles in fact acquire exponentially small imaginary parts.
These poles are the highly stable poles just below the top of the potential barrier and they connect onto the poles (\ref{eq:twofinal}) of case II that we discussed in the main text. To make the connection precise one should map from $\e$ to $\hat \w$ using $\e = \g A/B$ are the expressions for $A$ and $B$ in the main text. That these poles are exponentially close to the real axis justifies the assumption above that $I_W$ and $T_W$ are real.

A second set of poles is obtained with $\e_1 > 0$. These poles are forced by (\ref{eq:polesexp}) to have an imaginary part. For simplicity, let us look for poles with $\e_1 \gg \e_2$. The real part of the poles are again given by (\ref{reale}) but now there is an imaginary part, from solving the real part of equation (\ref{eq:polesexp}) given by
\be
\e_2 = \frac{ \e_1 \pi}{\log{|\e_1|}-1 - 2 T_W} \,.
\ee
We will shortly show that $T_W$ is large in the regime we are interested in ($1 \ll |\e| \ll \gamma$), and therefore our assumption $\e_1 \gg \e_2$ is consistent. In fact, we will find
\be\label{eq:TI}
I_W \sim \sqrt{g_\infty} \sqrt{h_\infty^2 -\hat m^2} \log \frac{1}{\hat \w^{1/z}}  \,, \qquad T_W \sim \log \sqrt{\g} \,.
\ee
Thus putting the two sets of poles together, the real and imaginary parts are
\bea
\e_1 & = & - \frac{2 \g \sqrt{g_\infty} \sqrt{h_\infty^2 -\hat m^2}}{z \log \g} \log \frac{1}{\hat \w} - \frac{\pi (2n + 1)}{\log \g}\,, \label{eq:spacing}\\
 \e_2 & = & \left\{ \begin{array}{ll}
0 & \textrm{if  $\e_1<0$}\\ 
-\frac{\e_1 \pi}{\log{\gamma}}& \textrm{if $\e_1>0$}
\end{array} \right. \,.
\eea
The geometry of these poles is discussed in the main text around (\ref{eq:spacingtext}).

Finally, let us derive the expressions (\ref{eq:TI}) for $T_W$ and $I_W$. Consider $I_W$ first. The upper endpoint of the integral expression  (\ref{eq:iw}) is inside the near horizon Lifshitz region. We can split the integral into a part that is contained in the Lifshtiz region and a part that is not. Schematically
\be
I_W = \int_1^{r_o} \sqrt{- \Delta V_\text{Lif.}} + \int_{r_1}^1 \sqrt{- \Delta V} \,.
\ee
The precise radius where we choose to make the split will not be important. We are aiming to isolating a contribution to the integral that is singular in the limit $\hat \w \to 0$. Away from the Lifshitz region, the potential does not have any singular $\hat \w$ dependence. Focussing on the near horizon contribution we can change variables to $x=r^z \hat \omega$ and obtain, using (\ref{eq:nearE}),
\be
I_W = \frac{\sqrt{g_\infty}}{z} \int_\omega^{x_o} \frac{dx}{x} \sqrt{\left((h_\infty + x)^2 - \hat m^2 \right) \left(1 - (x/x_o)^{2/z}\right)} \,.
\ee
Recall that $x_o$ is given by (\ref{eq:xo}) and does not depend on $\hat \w$. The upper limit of the integral is therefore regular. The lower limit leads to a logarithmic divergence with $\hat \w$ that we have quoted in (\ref{eq:TI}). The precise radius at which we evaluate the lower limit appears inside the logarithm and is therefore subleading as $\hat \w \to 0$.

For $T_W$ we can also perform the same change of variables in (\ref{eq:tw}). The integral part of $T_W$ is regular near the upper limit $r_o$ by construction, so the same reasoning applies and we don't need to worry about this contribution. The lower limit gives us a finite contribution. This contribution is subleading with respect to the extra piece of $T_W$, outside the integral, in the large $\gamma$ limit. In this terms we see that the $\omega$ contributions cancel and there is no frequency dependence
\be
T_W \sim  \log{\sqrt{\gamma B}|r_o - r_1|} \sim \log{ \sqrt{\gamma B} \left(x_o/\omega\right)^{1/z}} \sim \log{\sqrt{\gamma }} \,.
\ee
This is the expression quoted in (\ref{eq:TI}).

\section{The WKB limit for small stars}
\label{sec:ee}

In this appendix we comment briefly on the validity of the WKB limit for small stars. The potential issue is that while $\g$ is being taken large as always, we are also taking the parameter $\lambda$ of (\ref{eq:lambda}) to be small.

Firstly recall from footnote \ref{foot:a} that in terms of the quantity $\hat \b$ of \cite{Hartnoll:2010gu} we have $\l \sim \hat \b^{-1/3}$. Furthermore, recall from \cite{Hartnoll:2010gu} that $\hat \b \sim e^2 \g^2$. It follows that $\bar \g \lambda \sim \g \lambda \sim \g^{1/3} e^{-2/3} \gg 1$, because we are taking $\g \gg 1$ while the electron star solution requires $e \ll 1$. Thus the regime of validity of out computations always has $\bar \g \lambda \gg 1$. We will see that this is sufficient to force us into the WKB regime even in the small star limit (and, in particular, is compatible with the WKB limit).

The most dangerous place in the potential in the small star limit is the boundary region between cases II and III that we considered in section \ref{sec:wktext} in the main text. Recall from footnote \ref{foot:ha} that a condition for WKB matching to be possible was that $\sqrt{\g B} (x_o/\hat \w)^{\frac{1}{z}} \gg 1$. Let us check whether this condition is compatible with the small star limit. Using the formulae (\ref{eq:BBB}) and (\ref{eq:xo}) for $B$ and $x_o$ respectively, and expanding for $\l \ll 1$, we find that
\be
\sqrt{\g B} (x_o/\hat \w)^{\frac{1}{z}} \sim
\left\{
\begin{array}{lr}
\bar \g^{1/2} & \qquad z < 2 \\
(\bar \g \lambda)^{1/2} & \qquad z \geq 2
\end{array}
\right. \,.
\ee
The discussion in the previous paragraph showed that $\bar \g \lambda \gg 1$ remained large even while $\lambda$ was taken small. This is then sufficient for the WKB condition to hold. Thus we find that the small star limit of electron stars is compatible with our WKB analysis.

Factors of $\lambda \ll 1$ appear in various physical quantities in the small star limit. The critical dispersion relation (\ref{eq:disper}) becomes $\hat \w \sim \lambda^{2-z} \hat k^z$ for $z \geq 2$. Also for $z \geq 2$ all the factors of $\g$ in the locations of the Green's function poles in (\ref{eq:spacingtext}) and (\ref{eq:ppp}) become replaced by $\g \lambda$. This allows one to tune the spacing between the poles independently of the scale set by $\g$.

It is interesting to calculate $I_W$ and $T_W$ in the small star limit for $z=2$. Here we can obtain exact expressions that can be matched against (\ref{eq:TI}) where we extracted the dominant contributions. We can use expressions (\ref{eq:IW}) and (\ref{eq:TW}) and plug in the exact WKB potential $\Delta V$ at $z=2$ and $k$ obeying the (exact) critical dispersion relation $\omega = \frac{k^2}{\sqrt{2} + 2 \lambda}$:
\be
\Delta V =  - g_{\infty}\frac{\omega^2}{r^2} \left(\frac{\lambda}{\omega} - r^2\right)^2 \,.
\ee
In the small $\omega$ limit we can perform the integrals exactly and obtain:
\be
I_W = \sqrt{g_{\infty}}\lambda \log \sqrt{\frac{\lambda}{\omega}} \quad \quad T_W = \log \sqrt{\gamma \lambda} \,,
\ee
as predicted by (\ref{eq:TI}), including the extra factors of $\lambda$ that we mentioned in the previous paragraph.

\end{document}